\documentclass[amssymb,aps,prd,nofootinbib,two column]{revtex4-2}

\usepackage{graphicx}
\usepackage{subcaption} 
\usepackage{mathtools}
\usepackage{url}
\usepackage[normalem]{ulem}
\usepackage{caption}
\usepackage{enumitem}

\usepackage[colorlinks=true, linkcolor=blue, citecolor=red, urlcolor=blue]{hyperref}

\newcommand{\nn}{\nonumber}
\newcommand{\be}{\begin{equation}}
\newcommand{\ee}{\end{equation}}
\newcommand{\bea}{\begin{eqnarray}}
\newcommand{\eea}{\end{eqnarray}}

\newcommand{\om}{\omega}
\newcommand{\del}{\partial}
\newcommand{\al}{\alpha}

\newcommand{\orcid}[1]{\href{https://orcid.org/#1}{\includegraphics[width=8pt]
{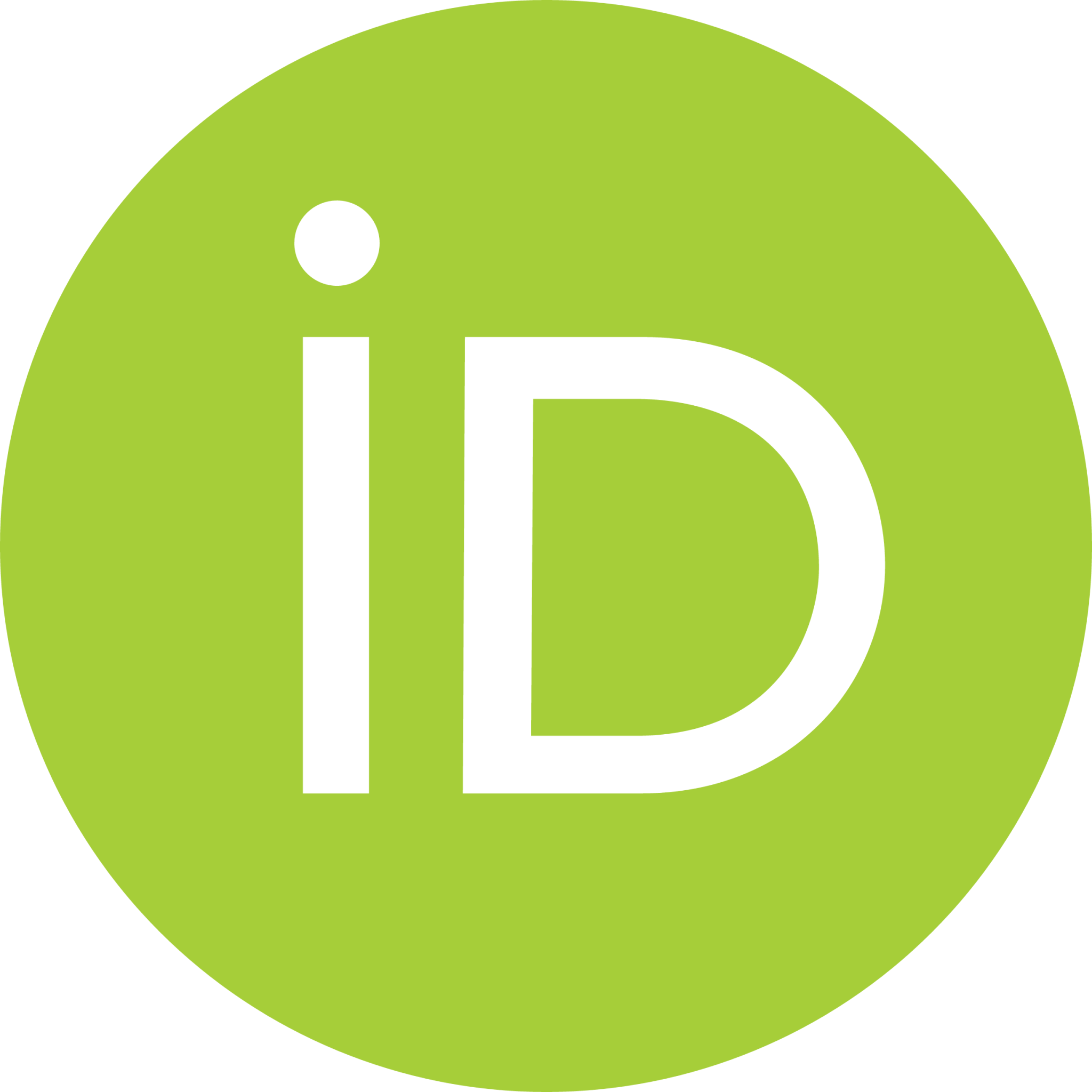}}}
\begin{document}
\title{Probing Dynamical Electrical Conductivity via Dilepton Emission: A Kinetic theory approach}

\author{Ashutosh Dwibedi\orcid{0009-0004-1568-2806}$^1$, Ankit Kumar Panda\orcid{0000-0002-9394-6094}$^{1}$, Sabyasachi Ghosh\orcid{0000-0003-1212-824X}$^1$, and Victor Roy\orcid{0000-0002-3741-9249}$^{2}$}
\affiliation{$^{1}$ Department of Physics, Indian Institute of Technology Bhilai, Kutelabhata, Durg 491001, India}
\affiliation{$^{2}$ School of Physical Sciences, National Institute of Science Education and Research, An OCC of Homi Bhabha National Institute, Jatni-752050, India}

\begin{abstract}
 Dileptons serve as a clean and penetrating probe of the Quark--Gluon Plasma created in high-energy heavy-ion collisions. In this work, we investigate thermal dilepton spectra and their elliptic flow through the dynamical conductivity that governs the production rate. The conductivity is obtained from the trace of the spectral function within relativistic kinetic theory using the Relaxation Time Approximation. This allows us to derive for the first time an analytical expression for the dilepton rate with explicit dependence on the relaxation time of quark–antiquark interactions. We find a non-monotonic dependence of the dilepton rate on the relaxation time and compare the resulting transverse momentum, invariant mass spectra and elliptic flow with previous quantum field theory results. The spectra and elliptic flow are obtained by integrating the rate over the full spacetime volume of the evolving medium, using temperature and flow profiles from realistic MUSIC hydrodynamic simulations without considering the effect of magnetic fields in the profiles itself. However, we study the role of an external space-time dependent magnetic field by making the conductivity anisotropic. At small relaxation times, magnetic fields have negligible impact, while for larger relaxation times and stronger initial fields, modifications of up to $\sim$20\% appear in both spectra and elliptic flow. Assuming instead a constant magnetic field of $\sim 1\,m_{\pi}^2$ at large relaxation times yields more modest effects, with changes of about 10\% in spectra and 5\% in elliptic flow. 
\end{abstract}
    
\maketitle
\section{Introduction}
The search for hot and dense QCD matter, popularly called Quark-Gluon Plasma (QGP), has been a central goal at facilities such as RHIC and the LHC. Direct confirmation of the QGP is challenging due to its extremely short lifetime~\cite{Kasza:2018qah}. As a result, its existence and properties are inferred through various indirect probes~\cite{Harris:2023tti,Rapp:2000pe}. Among the most prominent of these are measurements of the transverse momentum spectra and the azimuthal anisotropies of the emitted particles, often quantified through flow harmonics~\cite{Poskanzer:1998yz,Heinz:2013th,Romatschke:2007mq}. One such probe is dilepton production~\cite{Shuryak:1978ij,Wolf:1990ur,Li:1995qm,Rapp:1999ej,Alam:1999sc,Rapp:2000pe,Chatterjee:2007xk,PHENIX:2009gyd,Chatterjee:2009rs,Mohanty:2011fp,Bhatt:2011kx,Rapp:2013nxa,STAR:2013pwb,Gale:2014dfa,Linnyk:2015rco,Ryblewski:2015hea,Zhou:2024yyo,Massen:2024pnj,Wu:2024pba,Wei:2024lah,Garcia-Montero:2024lbl,Nishimura:2023not,Churchill:2023vpt,Coquet:2023wjk,Ogrodnik:2023qzw,Nishimura:2023oqn,Speranza:2018osi,Naik:2025pjt}. Due to their minimal final-state interactions with the QCD medium, dileptons provide a clearer signal, offering unique insight into the space-time evolution of the medium. They are produced throughout the space-time evolution of collision from the early partonic stage~\cite{PhysRevLett.64.2242} to the late hadronic stage~\cite{Rapp:1997fs,Rapp:2000pe,Li:1998ma,Ghosh:2010wt,Ghosh:2011gs}. Their production includes both thermal and non-thermal sources~\cite{Rapp:2000pe,Rapp:2013nxa}. In the early stages of the collision, non-thermal contributions primarily stem from Drell-Yan annihilation processes~\cite{Drell:1970wh}, which generate lepton pairs. At later times, following freeze-out, additional non-thermal dileptons are produced through the electromagnetic decays of long-lived hadrons—such as Dalitz decays (e.g., $\pi^{0}$ or $\eta \rightarrow \gamma l^{+}l^{-}$) and direct decays of vector mesons like the $\omega$ and $\phi$~\cite{Rapp:2013nxa}. Along with the above-mentioned processes, other sources are semi-leptonic decays of the correlated open charm or bottom pairs (like the decay of $\text{D}\bar{\text{D}}$ or $\text{B}\bar{\text{B}}$ pairs~\cite{PhysRevC.54.2606}). More details on the different contributions to the dilepton production can be found in~\cite{Rapp:2013nxa,Linnyk:2015rco}.
These non-thermal contributions provide an additional component beyond the thermally produced ones, which are not considered in the present study.
The sources of thermal production of the dileptons are however the in-medium $q\bar{q}$ annihilation in the partonic stage to the in-medium vector meson decays in the hadron gas stage. These thermal parts carry information about the produced quark and hadronic medium. It is well known that the low mass dilepton spectra ($0.5-1$ GeV) are dominated by the hadronic matter sources, and the intermediate mass dilepton spectra ($1.5-3$ GeV) mainly draw their contribution from the $q\bar{q}$ annihilation inside QGP. 

It is well known that in addition to the formation of the QGP medium, relativistic heavy-ion collisions (HICs) also generate extremely strong electromagnetic fields, arising from two possible ways--fast-moving spectator nucleons~\cite{Tuchin:2013apa,Tuchin:2013ie,Gursoy:2014aka,Deng:2012pc,Voronyuk:2011jd,Skokov:2009qp,Roy:2015coa,Gursoy:2018yai,Panda:2025lmd,Panda:2024ccj,Palni:2024wdy} and participant-induced fields particularly in scenarios involving charged expanding fluids and in presence of baryon stopping, as discussed in Refs.~\cite{Panda:2024ccj,Dash:2023kvr,Palni:2024wdy}.
 In the presence of electromagnetic fields, several studies have investigated bulk observables~\cite{Nakamura:2022ssn,Roy:2017yvg,Panda:2025lmd,Panda:2023akn,Parida:2025ddt,Palni:2024wdy} suggesting the splitting of flow harmonics arising from the inclusion of such fields.

Generally, in the study of thermal dilepton production, the dynamical conductivity, which characterizes the material's ability to transport the electric charges when subjected to time-varying fields of the medium plays a central role. In this work, we first establish the connection between the thermal dilepton rate and the dynamical conductivity, and then proceed to evaluate the latter for the medium under consideration. From a quantum field theoretical perspective, the thermal dilepton rate is directly proportional to the dynamical conductivity~\cite{Ding:2010ga,Wang:2022jxx,Wang:2021eud,Rapp:2024grb}. In the current study we compute the conductivity using the Boltzmann transport equation (BTE) within the relaxation time approximation (RTA). This approach naturally leads to a relaxation time-dependent expression for the thermal dilepton production rate. Moreover, as discussed earlier there is possibility of large magnetic fields in the reaction zone of non-central HICs which can affect the dilepton production rate. There have been several attempts in the literature to corroborate the effects of magnetic fields using the dilepton rate and spectrum~\cite{Bandyopadhyay:2016fyd,Sadooghi:2016jyf,Bandyopadhyay:2017raf,Ghosh:2018xhh,Islam:2018sog,Das:2019nzv,Hattori:2020htm,Ghosh:2020xwp,Chaudhuri:2021skc,Das:2021fma,Wang:2022jxx,Mondal:2023vzx,Castano-Yepes:2024vlj,Gao:2025prq,Panda:2025yxw}. Specifically, the Refs.~\cite{Bandyopadhyay:2016fyd,Sadooghi:2016jyf,Ghosh:2018xhh,Wang:2022jxx} contain the dilepton production rate from QFT calculated by evaluating the one-loop photon polarization tensor in the presence of strong magnetic fields. Whereas the authors of the Refs.~\cite{Bandyopadhyay:2017raf,Das:2019nzv} have calculated the rate in the weak field limit. The effective field theory models like NJL and PNJL have also been used to evaluate the dilepton rate in presence of magnetic fields~\cite{Islam:2018sog,Das:2021fma,Ghosh:2020xwp,Chaudhuri:2021skc,Mondal:2023vzx}. Unlike most of the previous works which consider a constant magnetic field the Ref.~\cite{Castano-Yepes:2024vlj} evaluated the rate in presence of variable magnetic fields within QFT. More recently, Gao \textit{et al.}~\cite{Gao:2025prq} have investigated the the effect of space-time varying magnetic field in the dilepton emission using (3+1) hydro simulations. A similar attempt using hydrodynamical simulation but with realistic magnetic field profile has also been done in~\cite{Panda:2025yxw}.
In the present study we incorporate the effect of such magnetic fields in the dilepton production rate by splitting the dynamical conductivity that enters rate in parallel and perpendicular directions. This anistropic nature of the conductivity results due to an additional timescale associated with cyclotron motion and have been studied earlier in the literature~\cite{Denicol:2019iyh,Ghosh:2019ubc,Panda:2021pvq,Panda:2020zhr,Bandyopadhyay:2023lvk,Satapathy:2021cjp,Das:2022lqh,Satapathy:2021wex,Dash:2020vxk,Dey:2020awu,Dey:2019vkn,Dey:2019axu}. We consider two simplified treatments of the magnetic field: one with a space-time-dependent profile and another with a constant field assumed over the entire space-time volume. The final analytical expression for the dilepton production rate, incorporating the effects of dynamical conductivity, is derived in the fluid rest frame.
 In this study, we aim to investigate the transverse momentum, invariant mass spectra, and elliptic flow ($v_2$) of dileptons. Additionally, we compare our results with previously obtained quantum field theory (QFT) predictions and examine the effects both with and without magnetic fields.
To this end, we compute these observables by integrating the local production rate over the full space-time volume in the laboratory frame. The transformation from the fluid rest frame to the lab frame is performed using realistic fluid velocity profiles generated from a $(2+1)$-dimensional hydrodynamic simulation using the \textsc{MUSIC} code~\cite{Schenke:2010nt,Schenke:2010rr,Paquet:2015lta,Huovinen:2012is}. The subtleties of the calculations and formalism will be discussed as and when required.

 In Sec.\eqref{formulation} we first outline the essential ingredients of the study, followed by a summary of the main steps taken to achieve our objectives. In Sec.~\eqref{Dilepton rate equation}, we discuss the thermal dilepton production rate and its relation to the dynamical electrical conductivity of the medium. The analytical expression for the dynamical conductivity  is derived in Sec.~\eqref{Conductivity calculation}. The numerical framework, including the \textsc{MUSIC} hydrodynamic simulation and the initialization procedure, is detailed in Sec.~\eqref{Numerical setup}. The results are presented and analyzed in Sec.~\eqref{Results}, followed by concluding remarks and potential directions for future work in Sec.~\eqref{Conclusions and Summary}.

{\bf Notations and Conventions:} Greek indices ($\mu, \nu, \alpha, \beta$, etc.) run from 0 to 3, and Latin indices ($j, k, l$, etc.) from 1 to 3, except $i$, reserved for $\sqrt{-1}$. Functions and contractions of four-vectors are written as $F(X^\mu) \equiv F(X)$ and $X^\mu Y_\mu \equiv X \cdot Y$; the squared magnitude as $X^2 \equiv X \cdot X$. Fourier transforms are defined by $\tilde{F}(q) \equiv \int d^4x e^{iq \cdot x} F(x)$ and $F(x) \equiv \int \frac{d^4q}{(2\pi)^4} e^{iq \cdot x} \tilde{F}(q)$. We use natural units with $\hbar = k_{\rm B} = c = \mu_0 = \epsilon_0 = 1$. Throughout this work, we use the metric with mostly negative signature, i.e., \( g_{\mu \nu} = \mathrm{diag}(+, -, -, -) \).

\section{Formulation}\label{formulation}
{Computation of the final dilepton spectra and the corresponding flow coefficients involves several key ingredients: an analytical calculation of the dilepton production rate, convolution of this rate with the hydrodynamic flow of the bulk medium, specification of the electromagnetic field, and formulation of the dynamical conductivity that enters the rate equation. Below, we outline these essential components step by step.}

\begin{itemize}[itemsep=2pt, parsep=0pt]
    \item We begin by calculating the dilepton production rate in the rest frame of the fluid. The central input to this rate is the {dynamical electrical conductivity} of the medium.

    \item The dynamical conductivity is computed using relativistic kinetic theory with the relaxation time approximation (RTA) as the collision kernel. This yields an expression that depends on temperature, invariant mass ($M$), frequency, and relaxation time ($\tau_c$).

    \item While the rest-frame rate provides valuable insight, our objective is to evaluate physically observable quantities such as the transverse momentum ($q_T$), invariant mass spectra and elliptic flow. To accomplish this, we incorporate model inputs from $(2+1)$-dimensional viscous hydrodynamic simulations using the \textsc{MUSIC} framework.

    \item The hydrodynamic profiles, obtained without including MHD evolution—which could, in principle, lead to different outcomes—provide the space-time dependent temperature and flow velocity fields. These fields are then used to boost the rest-frame rate to the laboratory frame and integrate it over the full medium evolution.

    \item Using the integrated rate, we evaluate the dilepton spectra and elliptic flow coefficients ($v_2$) both with and without magnetic field effects, introduced through the dynamical conductivity in the rate. This enables a detailed study of their dependence on relaxation time, temperature, and invariant mass.
\end{itemize}

\vspace{-20.5pt} 
\section{Dilepton rate equation}\label{Dilepton rate equation}
The rate of production of dileptons ($l^{+}l^{-}$) with total four momenta $q^{\mu}\equiv q_{1}^{\mu}+q_{2}^{\mu}\equiv(q^{0}=\omega,\vec{q})$ from the thermalized QGP can be expressed as~\cite{McLerran:1984ay},
\bea
\frac{d^{4}N}{d^{4}x~d^{4}q}=\frac{\al}{12 \pi^{4}}~\frac{L(q^{2})}{q^{2}} ~f_{\rm BE} (\om) ~g^{\mu\nu}~ {\rm Im} \Tilde{G}^{\rm R}_{\mu\nu} (\omega,\vec{q})~, \label{D1}
\eea 
where the phase-space factors for dileptons $L(q^{2})=\big(1+\frac{2m_{l}^{2}}{M^{2}}\big)\sqrt{1-\frac{4m_{l}^{2}}{M^{2}}}$ with $m_{l}$ and $M\equiv\sqrt{q^{2}}$ being the mass of the leptons and invariant mass of the produced dileptons. The $f_{\rm BE}$ and $\al=\frac{e^{2}}{4\pi}$ are respectively the Bose-Einstein distribution of the virtual photons ($\gamma^{*}$) and electromagnetic coupling constant. The retarded propagator $\Tilde{G}^{\rm R}_{\mu\nu}$ appearing in the expression~\eqref{D1} can be expressed as an electric current-current correlator~\cite{Ghosh:2010wt,Rapp:2000pe},
\bea
\tilde{G}^{\rm R}_{\mu\nu}(\omega,\vec{q})=-i\int d^{4}x~ e^{iq \cdot x}~\theta(x_{0})\langle[J_{\mu}(x),J_{\nu}(0)]\rangle_{\beta}~\label{new1},
\eea
where the retarded propagator in the position space reads $G^{\rm R}_{\mu\nu}(x-x^{\prime})=-i\theta(x_{0}-x_{0}^{\prime})~\langle[J_{\mu}(x),J_{\nu}(x^{\prime})]\rangle_{\beta}$. The $J_{\mu}$ appearing in the above expressions is the electric current of the quarks and $\langle\cdot\rangle_{\beta}$ stands for thermal averaging. Using the methods of thermal field theory one can calculate the trace of retarded propagator in leading order as ${\rm Im} (g^{\mu\nu}G^{\rm {R}}_{\mu\nu})=\frac{q^{2}e^{2}}{2\pi}$ (assuming massless quarks). Using Eq.~\eqref{D1} to leading order the dilepton production rate via the process $q\bar{q}\rightarrow \gamma^{*}\rightarrow l^{+}l^{-}$ can be expressed as~\cite{Sarkar:2012ty},
\bea
\frac{d^{4}N}{d^{4}x~d^{4}q}&=&\frac{\al^{2}}{6 \pi^{4}}~L(q^{2}) ~f_{\rm BE} (\om)\label{QFTGhosh}.
\eea

 We wish to express the dilepton emission rate in terms of the dynamical conductivity $\tilde{\sigma}(\om, \vec{q})$ of the QGP medium. This will be done by linking the retarded propagator with the dynamical conductivity. The terminology ``dynamical conductivty" is used to differentiate from the traditional conductivity which may be considered as static limit of the ``dynamical conductivity" owing to the Kubo's relation~\cite{Ghosh:2016yvt}. This dynamical conductivity $\tilde{\sigma}(\omega,\vec{q})$ is proportional to the spectral function of the electrical conductivity tensor whose $\omega\rightarrow 0,~\vec{q}\rightarrow 0$ gives us the tradiational electrical conductivity definition~\cite{Ghosh:2016yvt}.  
 For our purpose we first note down the induced electric current $J_{\mu}(x)$ as a response to an external electromagnetic four potential $A^{\mu}(x)$ as~\cite{Fernandez-Fraile:2009eug,Fernandez-Fraile:2005bew},   
\bea
J_{\mu}(x)=\int d^{4}x^{\prime}~ G^{\rm R}_{\mu \nu}(x-x^{\prime}) ~A^{\nu}(x^{\prime})~.\label{D2}
\eea
One can easily express the above relation in Fourier space as, $\tilde{J}_{\mu}(\om,\vec{q})=\Tilde{G}^{\rm R}_{\mu\nu}~\Tilde{A}^{\nu}(\om, \vec{q})$. To express the current $\tilde{J}_{\mu}$ in terms of electric field we go back to the definition of $A^{\mu}$ and perform the following steps,
\bea
&&\Tilde{A}^{\mu}(q) = \int d^{3}\vec{x}~ e^{-i\vec{q}\cdot \vec{x}} 
\int^{\infty}_{-\infty} e^{i\om t} A^{\mu}(t,\vec{x})~ dt , \nn \\
&& \implies \Tilde{A}^{j}(q) = \frac{1}{i\omega} \Tilde{E}^{j}(q)~.
\label{D3}
\eea
The last line follows from integration by parts on the time coordinate, choosing the gauge of vanishing $A^{0}$~\cite{Fernandez-Fraile:2009eug,Fernandez-Fraile:2005bew}, and using $\vec{E}=-\del_{t}\vec{A}$. We choose the specific gauge $\tilde{A}^{0}=0$ throughout the paper. Comparing Eq.~\eqref{D3} with the Fourier space version of Eq.~\eqref{D2} we get $\Tilde{\sigma}^{jk}\equiv\frac{\Tilde{G}^{\rm (R)j k}}{i \om}$. The temporal part of the retarded Green's function can be obtained by demanding its Gauge invariance\footnote{Under a Gauge transformation $A^{\prime\mu}= A^{\mu}+\del^{\mu}\chi,\implies \tilde{A}^{\prime\mu}= \tilde{A}^{\mu}-iq^{\mu}\tilde{\chi}$ and $\tilde{j}^{\mu}=\Tilde{G}^{\text{R}}_{\mu\nu} \tilde{A}^{\nu}=\Tilde{G}^{\text{R}}_{\mu\nu} \tilde{A}^{\prime\nu}$} which imply $\Tilde{G}^{\text{R}}_{\mu\nu}q^{\nu}=0$. The temporal part is then readily expressed as $\tilde{G}^{\rm R}_{00}=-\tilde{G}^{\rm R}_{0k}\frac{q^{k}}{\omega}$.
  Substituting this expression in Eq.~\eqref{D1} we get the dileptons rate equation in terms of dynamical conductivity and temporal component of Green's function as follows,
\bea
\frac{d^{4}N}{d^{4}x~d^{4}q}&=&\frac{\al}{12 \pi^{4}}~\frac{L(q^{2})}{q^{2}} ~f_{\rm BE} (\om) \left[ {\rm Im} \Tilde{G}_{k}^{{\rm (R)}k}+{\rm Im} \Tilde{G}^{{\rm (R)}0}_{0}\right]\nonumber\\
&=&\frac{\al}{12 \pi^{4}}~\frac{L(q^{2})}{q^{2}} ~f_{\rm BE} (\om)\nonumber\\
&&\left[\om~ {\rm Re} \Tilde{\sigma}_{k}^{k} (\omega,\vec{q})-\frac{q^{k}}{\omega}{\rm Im} \Tilde{G}^{{\rm (R)} 0}_{k}\right].\label{D4}
\eea
The above equation serves as the central equation in the present study, with particular emphasis on the trace of the retarded Green's function ($\tilde{G}^{\text{R}}_{\mu\nu}$), which will be evaluated using relativistic kinetic theory. In the following section, we proceed to compute the electrical conductivity and the temporal part of the $\tilde{G}^{\text{R}}_{\mu\nu}$ from the Boltzmann transport equation (BTE) similar to Refs.~\cite{Formanek:2021blc,Grayson:2022asf}.
\section{Conductivity calculation}\label{Conductivity calculation}
Let us start by defining the diffusion current from the relativistic kinetic theory,
\bea
\Tilde{J}^{\mu}=\sum _{j} g_{j} Q_{j}\int ~\frac{d^{3}\vec{p}_{j}}{(2\pi)^{3}} \frac{p^{\mu}_{j}}{p_{0j}}~ \Tilde{\delta f_{j}}~,\label{D5}
\eea
where $j$ is the particle index which runs over of the quarks ($u,d,s$ and their anti-particles), $Q_{j}$ are respective electric charges of the quarks, $g_{j}$ is the degeneracy factor (spin and color) and $\mu$ stands for the different components of the total electric current. The $\vec{p}_{j}$ and $p^{0}_{j}$ are the momentum and energy of the quarks, respectively. The $\Tilde{\delta f}_{j}$ are the off-equilibrium correction to the equilibrium distribution functions $f^{0}_{j}$ in response to the external driving force  $\vec{F}_{j}=Q_{j}\vec{E}$. The macroscopic expression for the 3-current as a response to the external electric field is $\Tilde{J}^{k}=\tilde{G}^{{\rm (R)}k}_{l}\tilde{A^{l}}=\Tilde{\sigma}^{k}_{l}~\Tilde{E}^{l}$, which is the Ohm's law in the Fourier space. By evaluating $\Tilde{J}^{k}$ from Eq.~\eqref{D5} and comparing it with the Ohm's law one can obtain the dynamical conductivity $\Tilde{\sigma}^{k}_{l}$. Similarly, $\tilde{G}^{{\rm (R)}0}_{k}$ can be evaluated using Eq.~\eqref{D5} and the relation $\tilde{J}^{0}=\delta\tilde{n}=\Tilde{G}^{{\rm (R)}0}_{k}\tilde{A}^{k}$.
 To fulfill our purpose we need to calculate the off-equilibrium correction $\Tilde{\delta f}_{j}$ as a linear response to the perturbing electric field. This can be done with the aid of BTE.
The BTE for the distribution $f_{j}$ of the quarks in the QGP can be written as,
\bea
\frac{\del f_{j}}{\del t} + \frac{\vec{p}_{j}}{p^{0}_{j}}\cdot \frac{\del f_{j}}{\del \vec{r}}+\vec{F}_{j}\cdot\frac{\del f_{j}}{\del \vec{p}_{j}}=C[f_{j}]~.\label{D6}
\eea
 $C[f_{j}]$ is the collision kernel arise due to random collision among the quarks, we will approximate it with the standard RTA, i.e.,  $C[f_{k}]=-\frac{u_{\mu}p_{j}^{\mu}}{p^{0}_{j}}\frac{f_{j}-f^{0}_{j}}{\tau_{c}}$ with $\tau_{c}\equiv\frac{1}{\Gamma}$ being the relaxation time of the quarks. The local equilibrium distribution functions in terms of fluid four velocity $u^{\mu}$ and medium temperature $T\equiv 1/\beta$ are given by $f^{0}_{j}=1/[e^{\beta u_{\mu}p^{\mu}_{j}}+1]$, where we assumed a vanishing chemical potential. Linearizing the BTE upto first order in $\delta f_{j}=f_{j}-f^{0}_{j}$ in the local rest frame $u^{\mu}=(1,\vec{0})$ we get:
\bea
\del_{t}~\delta f+\frac{\vec{p}}{p^{0}}\cdot \del_{\vec{r}}~\delta f +\frac{\delta f}{\tau_{c}}= -Q~ \vec{E}\cdot \del_{\vec{p}} f_{0},~\label{D7}
\eea 
where we ignored the quark labels which will be retained during the final evaluation of the dynamical conductivity. Taking Fourier transformation over the space-time coordinates in both sides of Eq.~\eqref{D7} we get the solution as, $
\Tilde{\delta f}(\vec{p},\om,\vec{q})=\frac{\vec{\Tilde{E}}\cdot \vec{v}_{p}}{i(\vec{q}\cdot\vec{v}_{p}-\om)+\Gamma}~Q\beta f_{0}(1-f_{0}),
$
where velocity of the quarks $\vec{v}_{p}=\vec{p}/p^{0}$. Substituting $\Tilde{\delta f}$ in the Eq.~(\ref{D5}) we get the charge and current density for a particular quark species to be:
\bea
&&\delta \tilde{n}=-g ~Q^{2} \beta  \int \frac{d^{3}\vec{p}}{(2\pi)^{3}} \frac{ (v_{p})_{k}}{i(\vec{q}\cdot\vec{v}_{p}-\om)+\Gamma} f_{0}(1-f_{0})~ i\omega\Tilde{A}^{k},\nonumber\\
&&\Tilde{J}^{k}=-g ~Q^{2} \beta  \int \frac{d^{3}\vec{p}}{(2\pi)^{3}} \frac{ (v_{p})_{l}~(v_{p})^{k}}{i(\vec{q}\cdot\vec{v}_{p}-\om)+\Gamma} f_{0}(1-f_{0})~ \Tilde{E}^{l}.\nn\\~\label{D9}
\eea
Comparing with the Ohm's law we obtain the conductivity and $\Tilde{G}^{{\rm (R)}0}_{k}$  as~\footnote{Note that the minus sign in conductivity appears because of the covariant component of quark velocity $(v_{p})_{l}$.}:
\bea
&&\Tilde{G}^{{\rm (R)}0}_{k}(\om, \vec{q})=-g~Q^{2} \beta  \int \frac{d^{3}\vec{p}}{(2\pi)^{3}} \frac{ i\omega~(v_{p})_{k}f_{0}(1-f_{0})}{i(\vec{q}\cdot\vec{v}_{p}-\om)+\Gamma},\nn\\
&&\Tilde{\sigma}^{k}_{l}(\om, \vec{q})=-g~Q^{2} \beta  \int \frac{d^{3}\vec{p}}{(2\pi)^{3}} \frac{ (v_{p})_{l}~(v_{p})^{k}f_{0}(1-f_{0})}{i(\vec{q}\cdot\vec{v}_{p}-\om)+\Gamma}.\nn~\\
\label{D10}
\eea
Retaining the particle index the total conductivity $\Tilde{\sigma}^{k}_{l}(\om, \vec{q}) \equiv \sum_{f} \Tilde{\sigma}^{k}_{l(f)}$ and $\Tilde{G}^{{\rm (R)}0}_{k}(\om, \vec{q})\equiv\sum_{f}\Tilde{G}^{{\rm (R)}0}_{k(f)}$ can be written as:
\bea
&&\Tilde{G}^{{\rm (R)}0}_{k}(\om, \vec{q})=-2g\sum_{f}Q_{f}^{2} \beta  \int \frac{d^{3}\vec{p}}{(2\pi)^{3}} \frac{ i\omega~(v_{p})_{k}f_{0}(1-f_{0})}{i(\vec{q}\cdot\vec{v}_{p}-\om)+\Gamma},\nn\\
&&\Tilde{\sigma}^{k}_{l}(\om, \vec{q}) = -2g \sum_{f} Q_{f}^{2} \beta \int \frac{ d^{3}\vec{p}}{(2\pi)^{3}} 
\frac{(v_{p})_{l} (v_{p})^{k}f_{0} (1-f_{0})}{i(\vec{q} \cdot \vec{v}_{p} - \om) + \Gamma} ,\nn\\
\label{D11}
\eea
where $f$ (not to be confused with the Fermi-Dirac distribution of quarks $f_{0}$) now runs over the flavors $u,d$ and $s$. An extra factor of two appears in Eq.~(\ref{D11}) because each quark and anti-quark contribute equally at vanishing chemical potential. It is noteworthy to point out that the above expression of the conductivity reduces to the usual static or DC conductivity $(\sigma_{\rm DC})^{k}_{l}$ in the limit $\om$ and $\vec{q}$ tend to zero. Simplifying Eq.~(\ref{D11}) and then substituting $\Tilde{G}^{{\rm (R)}0}_{k}$ and $\Tilde{\sigma}^{k}_{l}$ in Eq.~(\ref{D4}) we get the dilepton production rate as:
\bea
\frac{d^{4}N}{d^{4}x~d^{4}q}
=\frac{\al}{12 \pi^{4}}~\frac{L(q^{2})}{q^{2}} ~f_{\rm BE} (\om)~3\sigma_{\rm DC} ~\mathcal{A}(\vec{q},\om,\tau_{c}),\label{D12}
\eea
where $\sigma_{\rm DC}=\sum_{f}\sigma_{{\rm DC}(f)}=(1/3)\sum_{f}Q_{f}^{2}\tau_{c}T^{2}=\frac{2e^{2}}{9}\tau_{c}T^{2}$ and $\mathcal{A}=\frac{1}{4|\vec{q}|\tau_{c}^{2}}\ln\frac{\tau_{c}^{2}(\omega+|\vec{q}|)^{2}+1}{\tau_{c}^{2}(\omega-|\vec{q}|)^{2}+1}$ with $e$ being the magnitude of the electron's charge. In arriving at Eq.~(\ref{D12}) we have assumed the quarks to be massless.
 
 Now we briefly discuss the possible change in the dilepton rate in presence of magnetic fields ($\vec{B}$). In the presence of magnetic fields the isotropic static conductivity of the medium becomes anisotropic. The magnitude of the conductivity component in the direction perpendicular to the magnetic field becomes proportional to an effective relaxation time $\tau_{f}^{\perp}=\tau_{c}/(1+(\tau_{c}/
 \tau_{B,f})^{2})$ that is a function of the usual relaxation time $\tau_{c}$ and the cyclotron time period $\tau_{B,f}=E/Q_{f}|\vec{B}|$. The thermal averaged effective relaxation time can be obtained by replacing the $\tau_{B,f}$ in $\tau_{f}^{\perp}$ with $\langle\tau_{B,f}\rangle\equiv\langle E\rangle/Q_{f}|\vec{B}|$, where $\langle E\rangle=\int f_{0}~ E d^{3}\vec{p}/\int f_{0}~ d^{3}\vec{p}$~. In the parallel direction to the magnetic field the conductivity remains same as of the isotropic conductivity. In the presence of magnetic fields we have $\sigma^{\perp,||}_{{\rm DC}(f)}=(1/3)Q_{f}^{2}\tau_{f}^{\perp,||}~T^{2}$, where $\tau_{f}^{||}=\tau_{c}$. Using the above analysis one can express the rate in Eq.~(\ref{D12}) in the presence of magnetic fields with the following replacement: $\sum_{f}3\sigma_{{\rm DC}(f)}\mathcal{A}\xrightarrow{}\sum_{f}2\sigma^{\perp}_{{\rm DC}(f)} \mathcal{A}_{f}^{\perp}$\\
 $+\sigma^{||}_{{\rm DC}(f)} \mathcal{A}_{f}^{||}$, where $\mathcal{A}_{f}^{\perp,||}=\frac{1}{4|\vec{q}|(\tau_{f}^{\perp,||})^{2}}\ln\frac{(\tau_{f}^{\perp,||})^{2}(\omega+|\vec{q}|)^{2}+1}{(\tau_{f}^{\perp,||})^{2}(\omega-|\vec{q}|)^{2}+1}$.
 Furthermore, the magnetic field profile used in this work assumes a Gaussian smearing of the sources in the transverse plane, encoded in a spatial profile function \(\rho(x, y)\). The magnetic field (specifically the \(y\)-component) at a fixed proper time \(\tau\) is modeled as:
\[
B_y(x, y, \tau) = \frac{B_0}{1 + \tau / {\tau}_d} \, \rho(x, y),
\]
where \(B_0\) is the peak magnetic field amplitude, \(\tau_d\) is the decay constant characterizing the field's decay time, and \(\rho(x, y)\) represents the transverse distribution of sources given by a Gaussian:
\[
\rho(x, y) = \exp\left[ -\frac{(x + b/2)^2 + (x - b/2)^2 + y^2}{2\sigma^2} \right],
\]
with \(b\) the impact parameter and \(\sigma\) controlling the width of the smearing. The profile of the same is shown below in ~Fig.\eqref{fig:By_contour}.
\begin{figure}[htbp]
    \centering
    \includegraphics[width=0.45\textwidth]{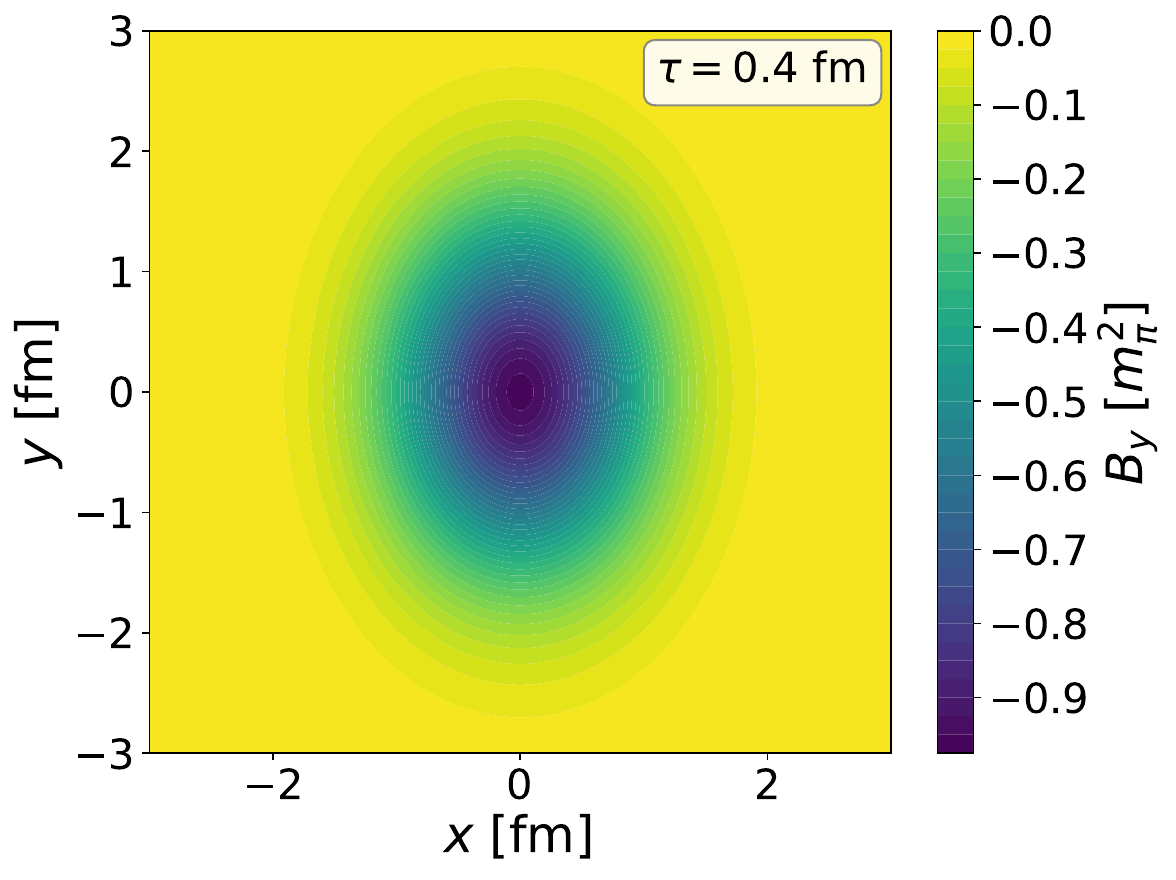}
    \caption{(Color online) Spatial distribution of the magnetic field component \(B_y(x, y)\) at fixed proper time \(\tau = 0.4\) fm and $\tau_d$ = 10 fm, with Gaussian smearing centered at the origin. The field is shown in units of \(m_\pi^2\).}
    \label{fig:By_contour}
\end{figure}
This form approximates the field generated by two colliding nuclei at early times in non-central HICs.

\section{Numerical setup}\label{Numerical setup}

In this study, we consider the collision of two heavy nuclei with their centers positioned symmetrically at \( x = \pm b/2 \), where \( b \) denotes the impact parameter and is oriented along the transverse \(x\)-direction. The colliding nuclei are taken to be gold (\(^{197}_{79}\mathrm{Au}\)), with atomic number \( Z = 79 \) and mass number 197. The center-of-mass energy per nucleon pair is fixed at \( \sqrt{s_{NN}} = 200 \)~GeV.

Using this setup, we investigate the formation and evolution of the QGP medium created in such high-energy HICs. The focus is on modeling the initial energy density profile generated in these collisions and studying its subsequent space-time evolution, which governs the temperature and fluid velocity profiles of the QGP. We begin by discussing the framework of relativistic hydrodynamic evolution used to describe the medium's dynamics, followed by introducing the initial condition employed to initialize the system at early times.

\subsection{Relativistic hydrodynamic evolution}

\begin{figure}
    \centering
    \includegraphics[width=0.9\linewidth]{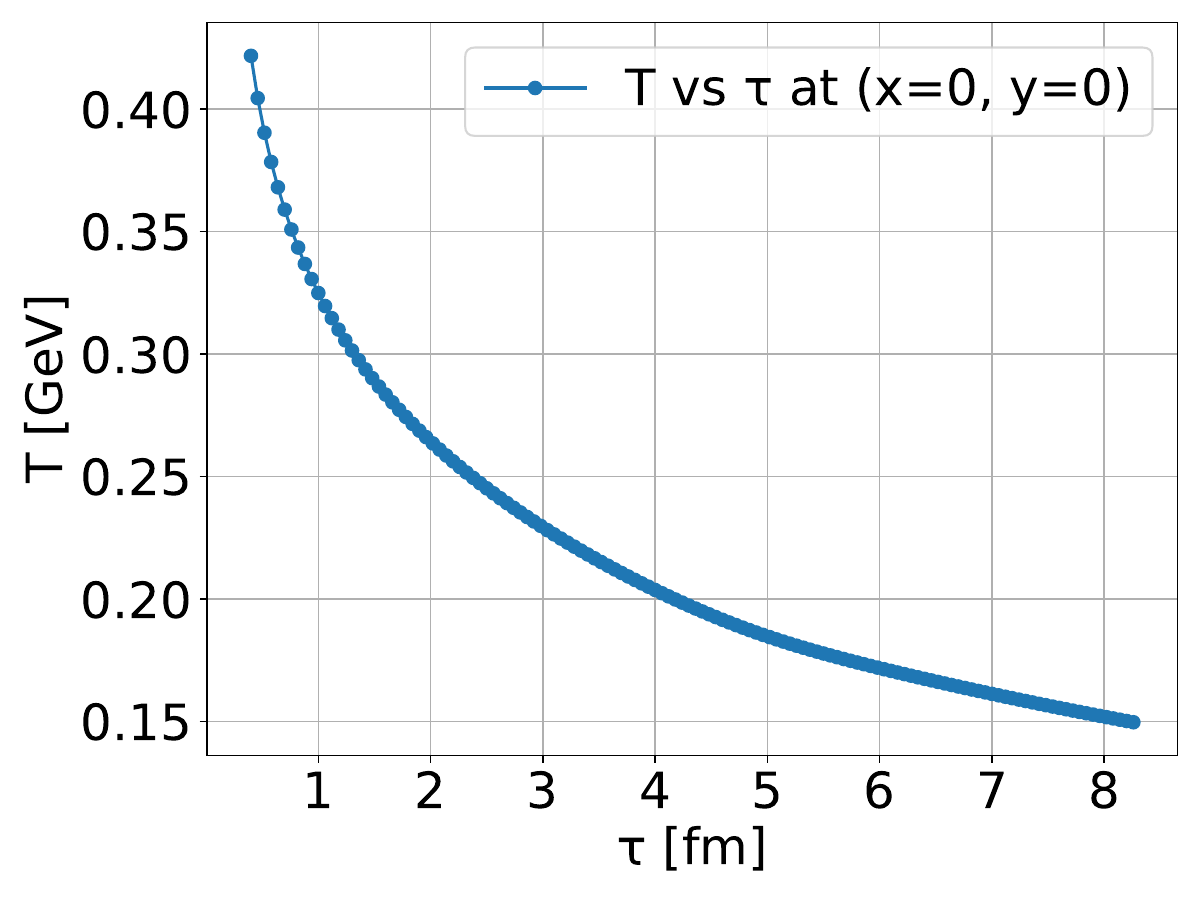}
    \caption{(Color online) The temperature evolution at the center of the collision zone for Au+Au collisions at \(\sqrt{s_{NN}} = 200~\text{GeV}\) and impact parameter \(b = 7~\text{fm}\), simulated using the 2+1D MUSIC hydrodynamic code with shear viscosity \(\eta/s = 0.08\) and zero bulk viscosity.
 }
    \label{fig:temperature}
\end{figure}

MUSIC is a {3+1 dimensional} viscous hydrodynamic framework capable of solving the full relativistic conservation equations for the energy-momentum tensor \( T^{\mu\nu} \) and net charge current \( N^\mu \). In this study, we employ its {2+1 dimensional boost-invariant} mode, which assumes longitudinal boost invariance and thus eliminates any dependence on spatial rapidity. 
\begin{figure}
	\centering
	\includegraphics[width=0.45\textwidth]{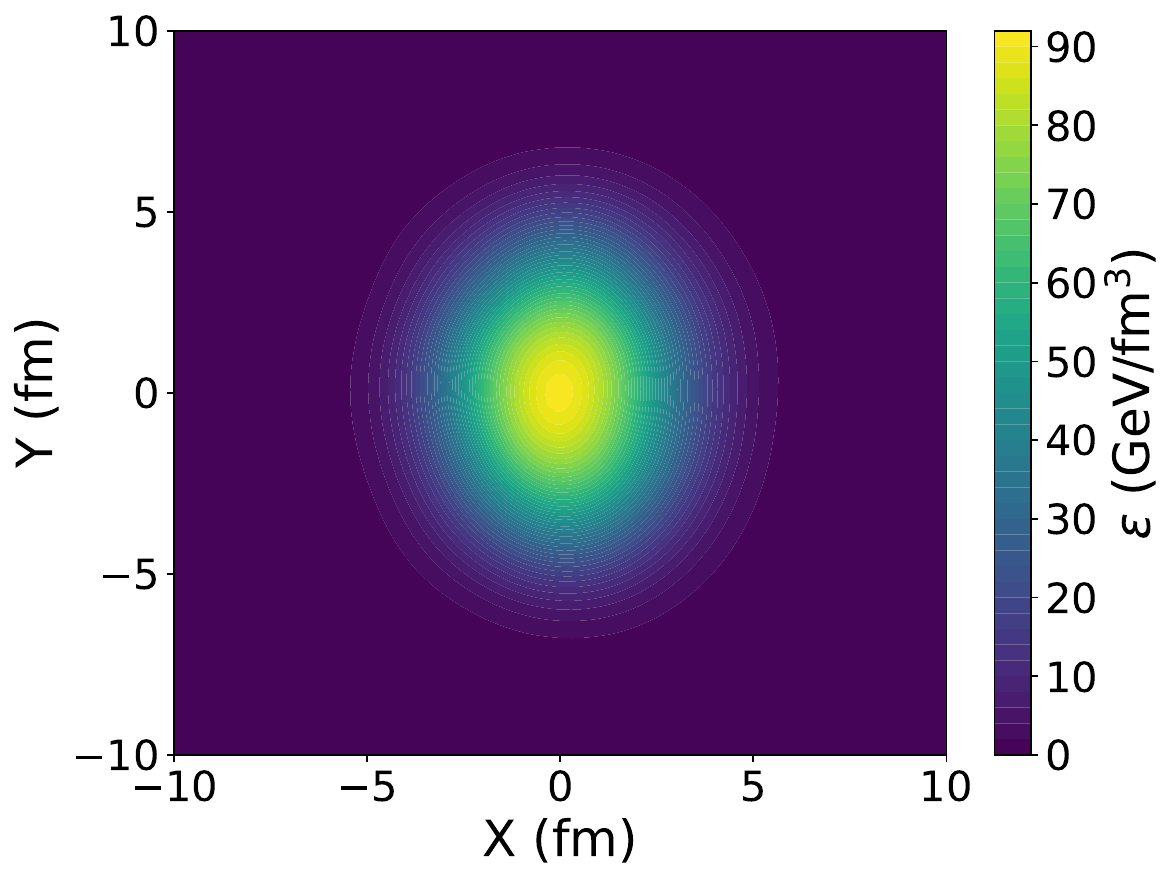}
	\caption{(Color online) Initial energy density profile for Au+Au collisions at \(\sqrt{s_{\mathrm{NN}}} = 200\) GeV with impact parameter \(b = 7\) fm.}
	\label{fig:initial_energy_density}
\end{figure}
The simulation is performed on a grid extending \(20\,\text{fm} \times 20\,\text{fm}\) in the transverse plane, discretized into \(200 \times 200\) spatial cells. The hydrodynamic evolution starts at a proper time of \( \tau_0 = 0.4\,\text{fm} \), with a time step of \( \Delta \tau = 0.06\,\text{fm} \) and proceeds up to a maximum time of \( \tau = 12\,\text{fm} \) with the spatial spacing of $\Delta x = \Delta y = 0.4\, \text{fm}$ .
Viscous effects are included through a constant shear viscosity to entropy density ratio \( \eta/s = 0.08 \), while bulk viscosity is neglected (\( \zeta/s = 0 \)).
The equation of state (EoS) used in the simulation is the lattice QCD EoS from hotQCD with UrQMD. To initiate the hydrodynamic evolution, a realistic initial energy density profile is required. In this work, we employ the well-established Monte Carlo Glauber model to generate the initial conditions using the usual procedure~\cite{Miller:2007ri}.
The initial energy density profile, averaged over 2000 events generated using the Monte Carlo Glauber model for Au+Au collisions at \(\sqrt{s_{\mathrm{NN}}} = 200\) GeV with an impact parameter \(b = 7\) fm is shown in Fig.~\eqref{fig:initial_energy_density}. This energy density profile is used as the initial condition for the hydrodynamic evolution within the boost-invariant 2+1D framework discussed above. After the hydrodynamic evolution, we obtain the evolution for the temperature as shown in Fig.~\eqref{fig:temperature}. 
%The evolution is done till all the fluid cell's energy density drops below a $T_{\text{min}}$ = 150 MeV.
\subsection{Calculation of spectra and flow harmonics}\label{spectra and flow}
The transverse momentum spectra of dileptons are calculated by integrating the differential production rate over the full spacetime volume of the evolving quark-gluon plasma. This is motivated by the fact that dileptons can be emitted continuously throughout the entire evolution of the medium.
In this study, the integration is performed only over those fluid cells with a local temperature above \(155\,\text{MeV}\) (inspired by the URQMD and Hot QCD EOS switching point~\cite{Moreland:2015dvc}), thereby restricting the contribution to regions of the system where the temperature is sufficiently high to support thermal dilepton production sourced only from QGP.
The total dilepton yield is thus obtained from the following expression:
\begin{equation}
\frac{dN}{d^{4}q} = \int d^{4}x \, \frac{d^{4}N}{d^{4}x \, d^{4}q} = \int \tau \, d\tau \, dx \, dy \, d\eta \; \frac{d^{4}N}{d^{4}x \, d^{4}q},
\end{equation}
where \( d^4x = \tau \, d\tau \, dx \, dy \, d\eta \) represents the differential element of the fluid volume in Milne coordinates. Again the four-momentum phase-space element can be written as: 
$
d^4q = q_T \, dq_T \, \, dM^2 \, dy \, d\phi_p$,
where \( q_T \) is the transverse momentum, \( M \) is the invariant mass of the dilepton pair, \( y \) is the rapidity, and \( \phi_p \) is the azimuthal angle of the dilepton momentum.
The anisotropic flow coefficient \( v_2 \), characterizing the second Fourier harmonic of the dilepton azimuthal distribution, is calculated as:
\begin{equation}
v_2(q_T,M) = \frac{\int d\phi_p \, \frac{dN}{q_T dq_T \, d\phi_p dM dy} \cos 2(\phi_p - \Psi_2 )}{\int d\phi_p \, \frac{dN}{q_T dq_T \, d\phi_p dM dy}}.
\end{equation}
Here, $\Psi_{2}$ is the second-order reaction plane angle, which sets the reference direction (relative to the $x$-axis in the lab frame) for measuring elliptic flow. The spectra and related observables are evaluated at mid-rapidity ($y = 0$), after integrating over the azimuthal angle $\phi_{p}$.

\section{Results}\label{Results}
With all the relevant expressions derived and the numerical setup in place, along with the observables defined, we are now prepared to explore the results. Our focus lies primarily on examining the variation of the dilepton yield with transverse momentum ($q_T$), invariant mass ($M$), relaxation time ($\tau_{c}$) and the corresponding elliptic flow as functions of $q_T$, $M$, and $\tau_{c}$ as outlined earlier. We also
\begin{figure}[htbp]
	\centering
	\includegraphics[width=0.9\linewidth]{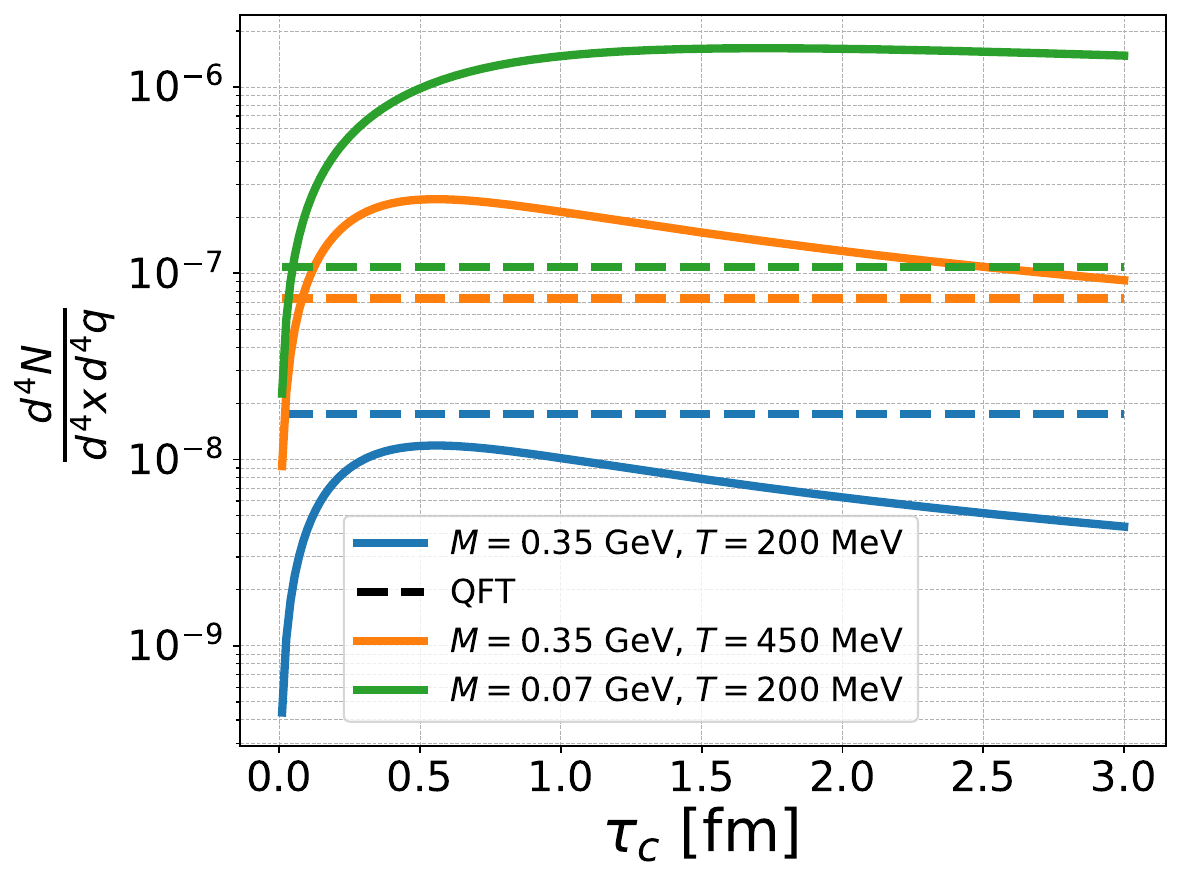}
	\caption{(Color online) Dilepton production rate $\left(\frac{d^4N}{d^4x\, d^4q}\right)$ is shown as a function of relaxation time $\tau_c$ at $M = 0.35~\text{GeV}$ and $q = 0.1~\text{GeV}$ for $T = 200$ and $450~\text{MeV}$, comparing kinetic theory with QFT. Results for $M = 0.07~\text{GeV}$ are also included.
	}
	\label{fig:dilepton_rate_tau_c}
\end{figure}
\begin{figure}[htbp]
	\centering
	\includegraphics[width=0.45\textwidth]{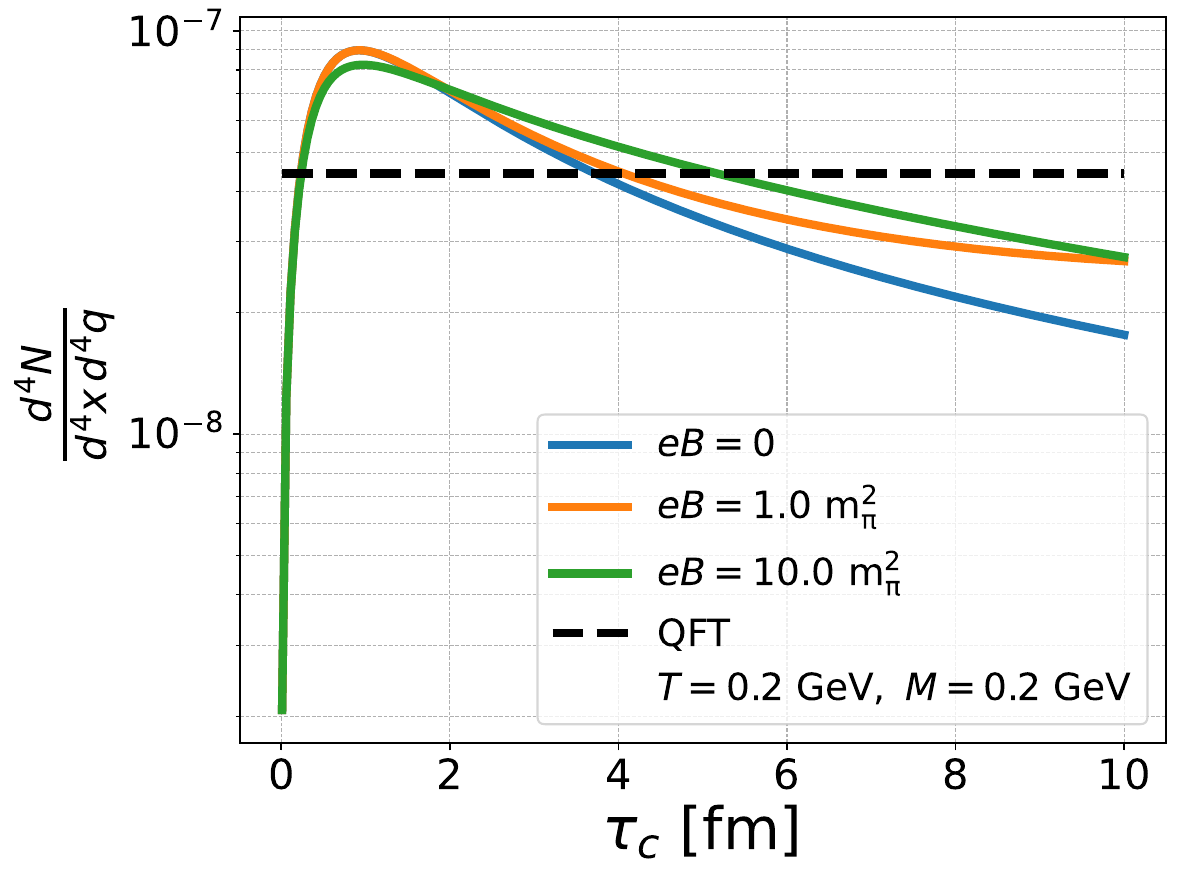}
	\caption{(Color online) Dilepton production rate as a function of relaxation time $\tau_c$ for $T = 0.2~\mathrm{GeV}$, q = 0.1 GeV and $M = 0.2~\mathrm{GeV}$. The blue curve corresponds to $eB = 0$ and the orange curve to $eB = 1~m_\pi^2$, green one for $eB = 10~m_\pi^2$ .}
	\label{fig:dilepton_rate_Bfield}
\end{figure}
\begin{figure*}[htbp]
	\centering
	\includegraphics[width=0.4\textwidth]{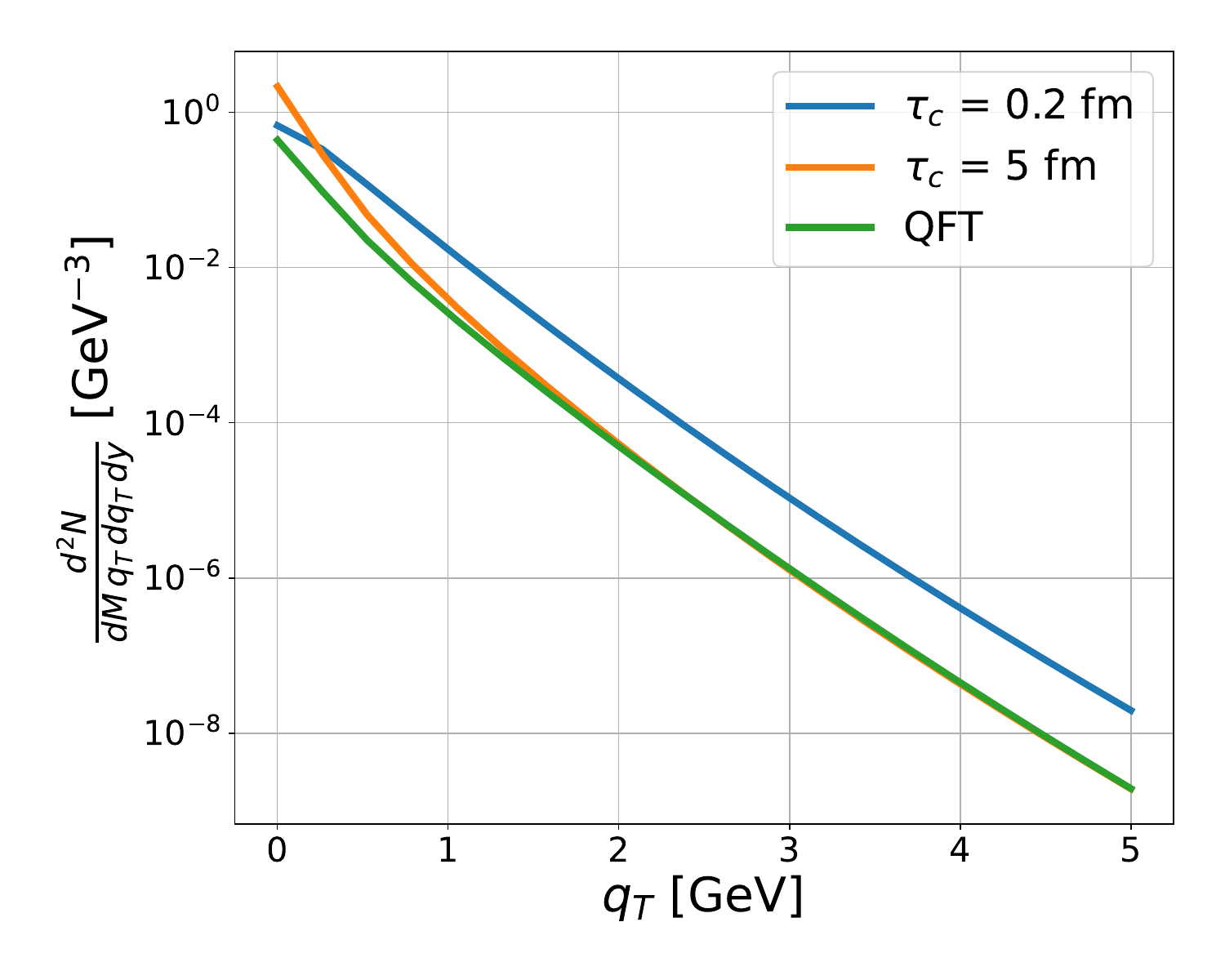}
	\includegraphics[width=0.4\textwidth]{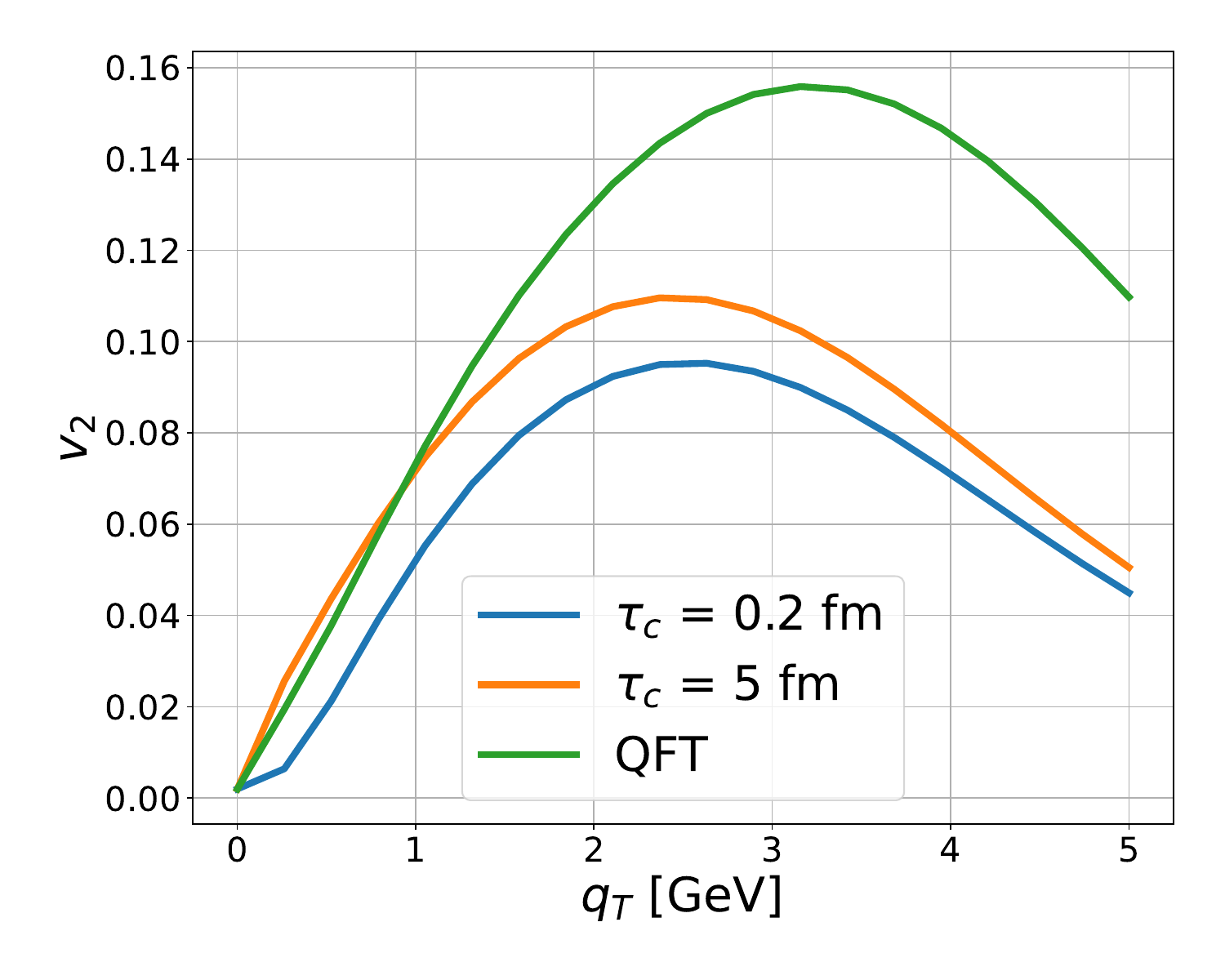}
	
	\vspace{-0.3cm} % Adjust spacing between rows if needed
	
	\includegraphics[width=0.4\textwidth]{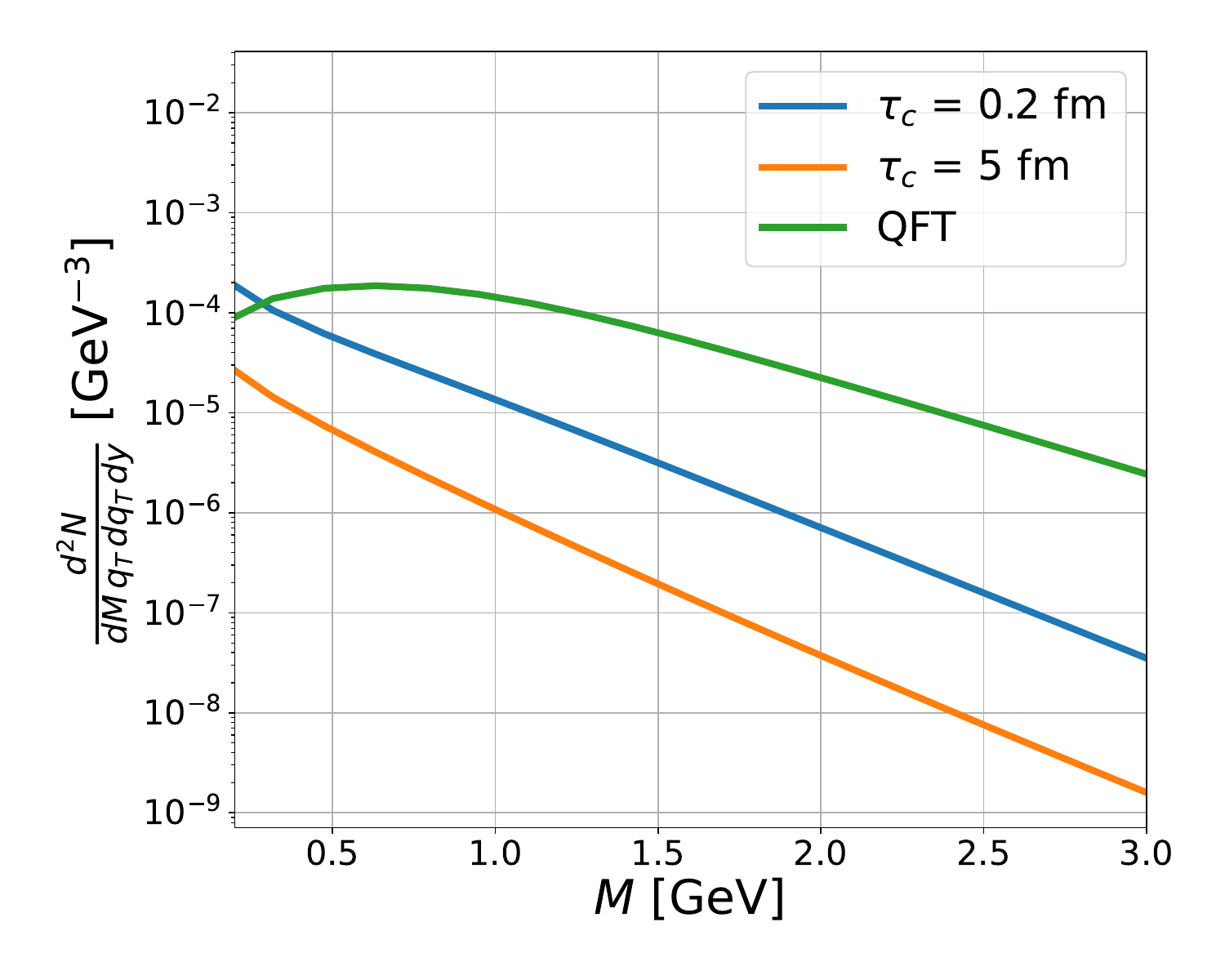}
	\includegraphics[width=0.4\textwidth]{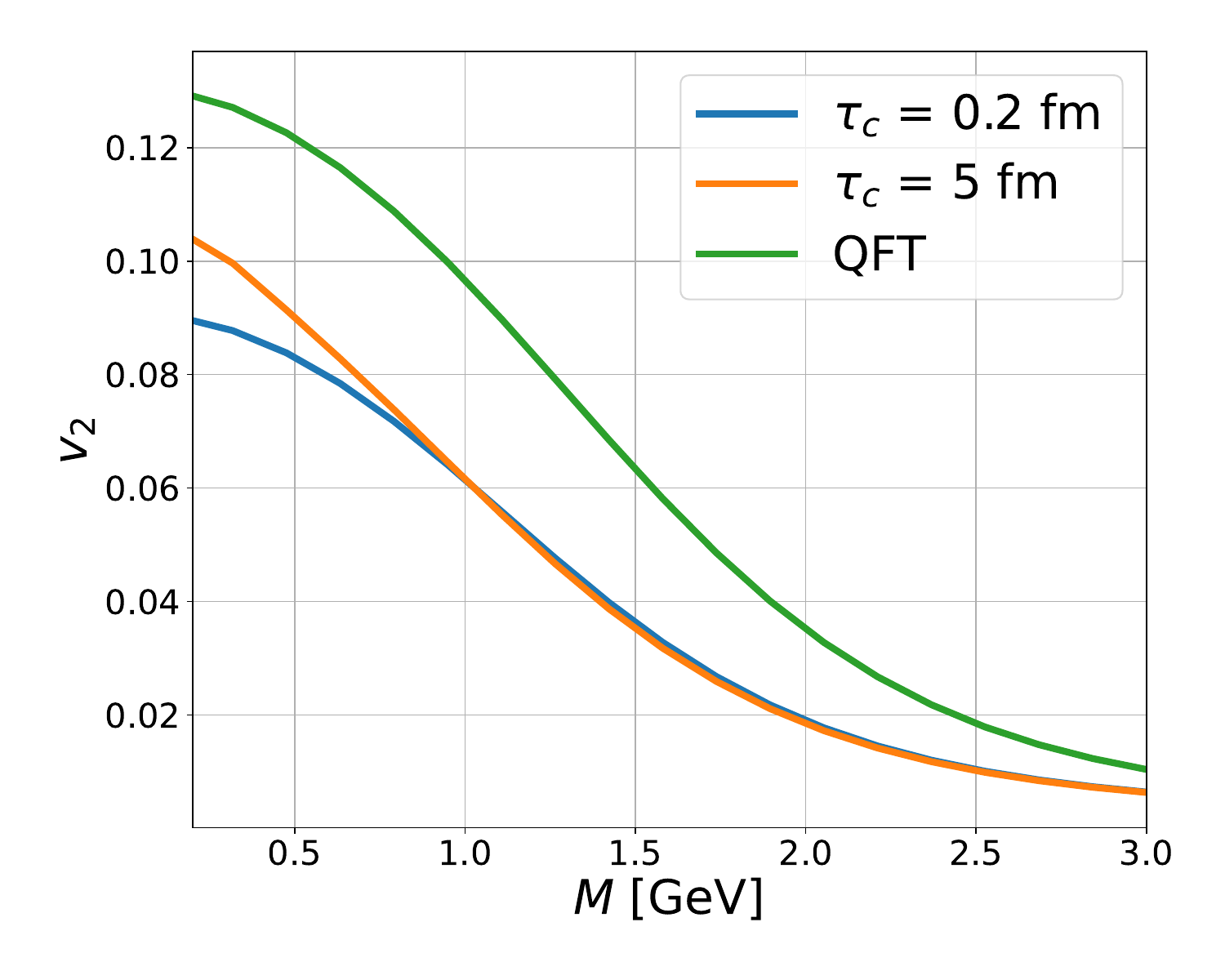}
	
	\caption{(Color online) Dilepton transverse momentum spectrum (top left panel) with the variation of elliptic flow coefficient 
		(top right panel) against $q_{T}$ for fixed \( M = 0.1\,\mathrm{GeV} \), and the invariant mass spectrum (bottom left panel) 
		with the variation of elliptic flow coefficient (bottom right panel) against $M$ for fixed \( q_T = 2\,\mathrm{GeV} \) at midrapidity (\( y=0 \)). 
		The spectrum and flow compare kinetic theory results for different relaxation times and QFT predictions (shown in green).}
	
	\label{spectra_v_2_noB}
\end{figure*}
 compare our results of rate, spectra and elliptic flow obtained from kinetic theory (KTh) (in Eq.\eqref{D12}) with that of QFT (in Eq.\eqref{QFTGhosh}). Let us begin by analyzing the profile of the dilepton production rate, which offers a static snapshot of the system, in contrast to the final spectra measured in the laboratory frame. It is also commonly called static dilepton rate.

 Fig.~(\ref{fig:dilepton_rate_tau_c}) illustrates the variation of the dilepton production rate (solid lines) with relaxation time \(\tau_c\) in the local rest frame with its QFT counterpart (dashed lines). The figure presents results for two different temperatures, \(T = 200\) MeV (blue curve) and \(T = 450\) MeV (orange curve), at a fixed invariant mass of \(M = 0.35\) GeV.
 \begin{figure*}
		 \centering
    \begin{subfigure}[t]{0.49\textwidth}
        \centering
        \includegraphics[width=0.82\linewidth]{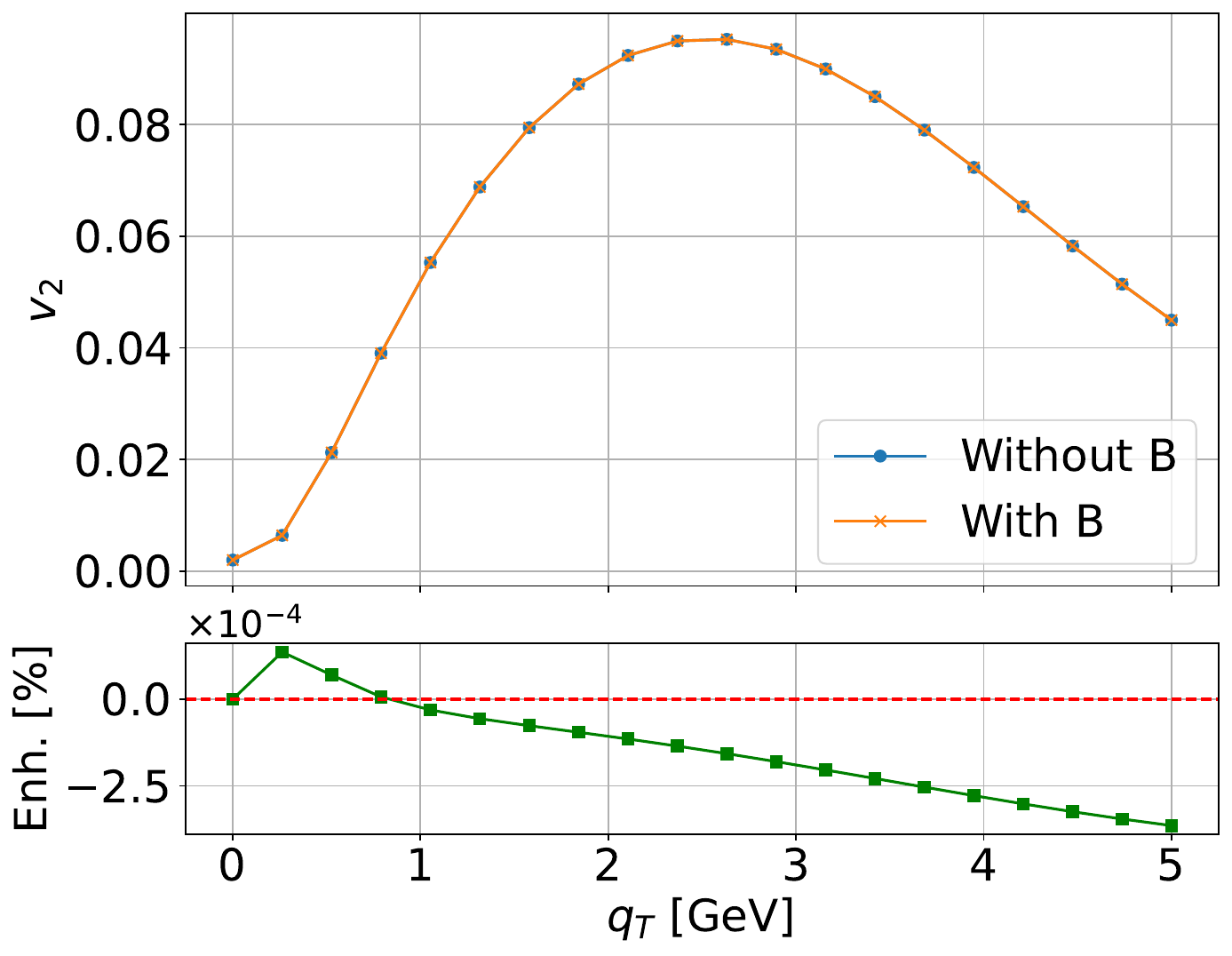}
        \label{subfig:v2_qT}
    \end{subfigure}
    \hfill
    \begin{subfigure}[t]{0.49\textwidth}
        \centering
        \includegraphics[width=0.82\linewidth]{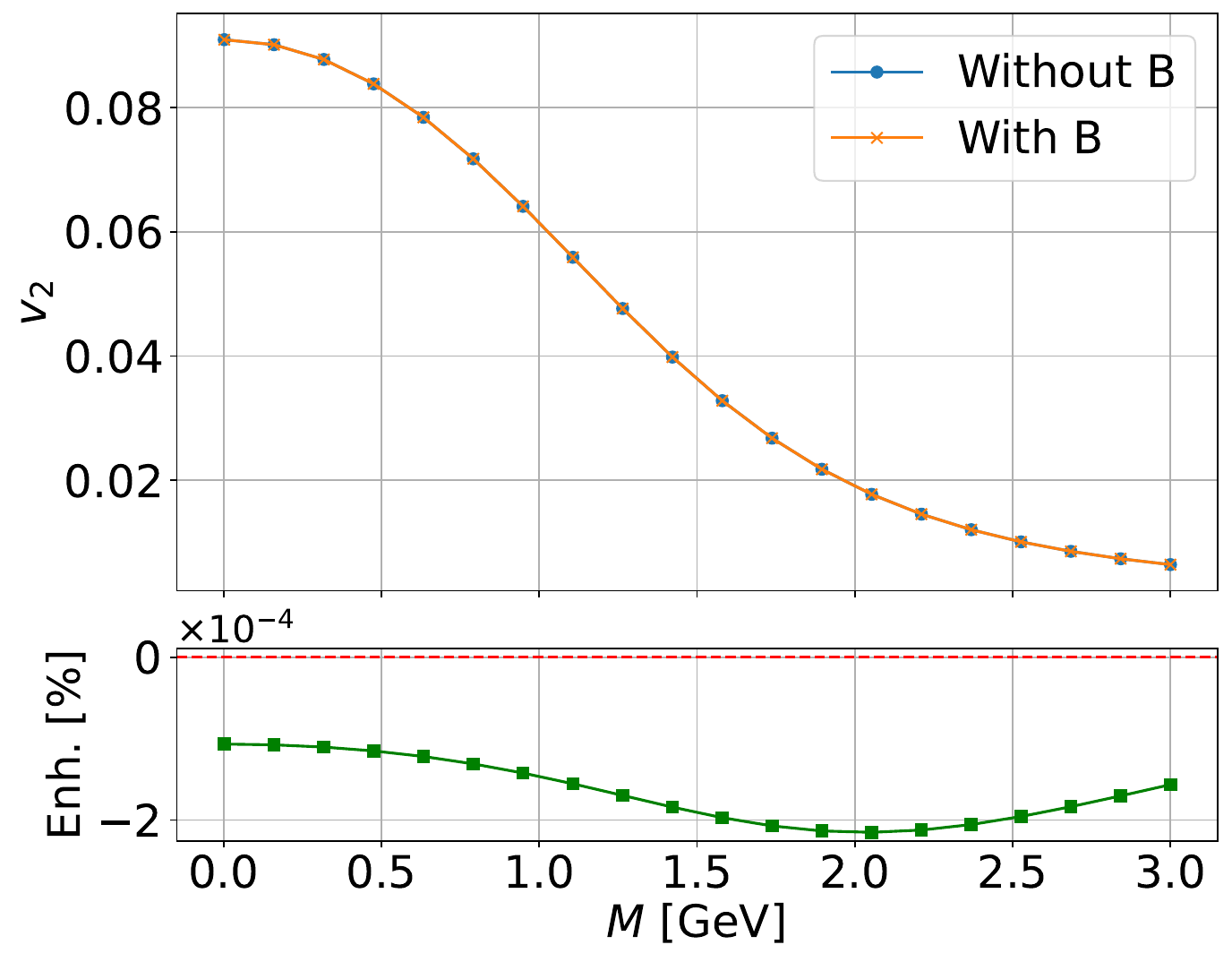}
        \label{subfig:v2_M}
    \end{subfigure}
		
		\caption{(Color online) Comparison of dilepton $v_2(q_{T})$ (left) at $M = 0.1$ GeV, and $v_{2}(M)$ (right) at $q_{T}=2$ GeV with and without $B$. 
			The relaxation time has been taken as $\tau_{c}=0.2$ fm. The lower panel shows the percentage enhancement 
			$(v_{2}(B)-v_{2}(B=0))\times 100$ / $v_{2}(B=0)$.}
		
		\label{fig:v2Bvary}
	\end{figure*}
 Additionally, the green curve shows the rate at \(T = 200\) MeV for a lower invariant mass of \(M = 0.07\) GeV.
The dilepton rate recorded in Eq.~\eqref{D12} initially increases with \(\tau_c\), reaches a peak, and then decreases as \(\tau_c\) continues to grow. This resonance-like behavior along with the peak in the dilepton rate is result of the factor $\frac{1}{\tau_{c}}\ln\frac{\tau_{c}^{2}(\omega+|\vec{q}|)^{2}+1}{\tau_{c}^{2}(\omega-|\vec{q}|)^{2}+1}$ that enters in the rate expression Eq.~\eqref{D12} via dynamical conductivity. In contrast, we see the constant lines for the QFT rates as they are relaxation time independent~\cite{Sarkar:2012ty}. It is also observed that the peak of the dilepton rate shifts to higher $\tau_c$ with a decrease in the invariant mass which also is an artifact of the factor mentioned above in our theory.
Moreover, the rate is highly sensitive to $M$ thus by reducing the mass from 0.35 GeV to 0.07 GeV (a factor of about five) results in an increase in the production rate by nearly two orders of magnitude for KTh and one order in magnitude for QFT. The extra increment in the KTh rate from QFT by reducing mass comes because of the extra factor (KTh/QFT) $\frac{\pi}{3}T^{2}\frac{1}{|\vec{q}|\tau_{c}M^{2}}\ln\frac{\tau_{c}^{2}(\omega+|\vec{q}|)^{2}+1}{\tau_{c}^{2}(\omega-|\vec{q}|)^{2}+1}$. Due to this extra factor, it is also interesting to note that above certain mass (around 0.3 GeV) the QFT results are dominant whereas below that the KTh results starts dominating for a particular value of $q$.
 Finally, as expected from thermal considerations, increasing the temperature leads to a substantial rise in the dilepton production rate, reflecting the enhanced thermal population of quarks and antiquarks at higher temperatures.

We now turn to the impact of magnetic fields on dilepton production. As shown in Fig.~\eqref{fig:dilepton_rate_Bfield}, the presence of a constant magnetic field—both at $1 m_{\pi}^{2}$ and $10 m_{\pi}^{2}$—clearly enhances the dilepton rate at fixed invariant mass $M$, magnitude of momentum $|\vec{q}|$, and temperature. This enhancement is particularly pronounced for large relaxation times ($\tau_{c} > 4$~fm) at lower field strengths, while for stronger fields (e.g., $10 m_{\pi}^{2}$) noticeable differences also emerge at smaller $\tau_{c}$ values. This is because the ratio $\tau_{c}/\tau_{B}$ becomes gradually non-negligible and dominant when one increases $\tau_{c}$ as well as magnetic field $B\propto \frac{1}{\tau_{B}}$. So, we notice that finite magnetic field can lift up dilepton rates but it is limited for large $\tau_{c}$ domain. Readers may notice that our present work first time addressed a KTh based dilepton rate expression which is quite sensitive with medium interaction information like relaxation time $\tau_{c}$. Large values of $\tau_{c}$ is assigned to weakly coupled QGP while small values of $\tau_{c}$ is associated with strongly coupled QGP. The magnitude of our KTh based dilepton rates for weakly and strongly coupled QGP medium are quite different. It is also evident from the static dilepton rate that weakly coupled QGP can be more sensitive to the magnetic field than the strongly coupled QGP. However, actual impact of magnetic fields on the dilepton production can be known after performing full evolution of the expanding QGP medium.

\begin{figure*}
	\centering
	\includegraphics[width=0.4\textwidth]{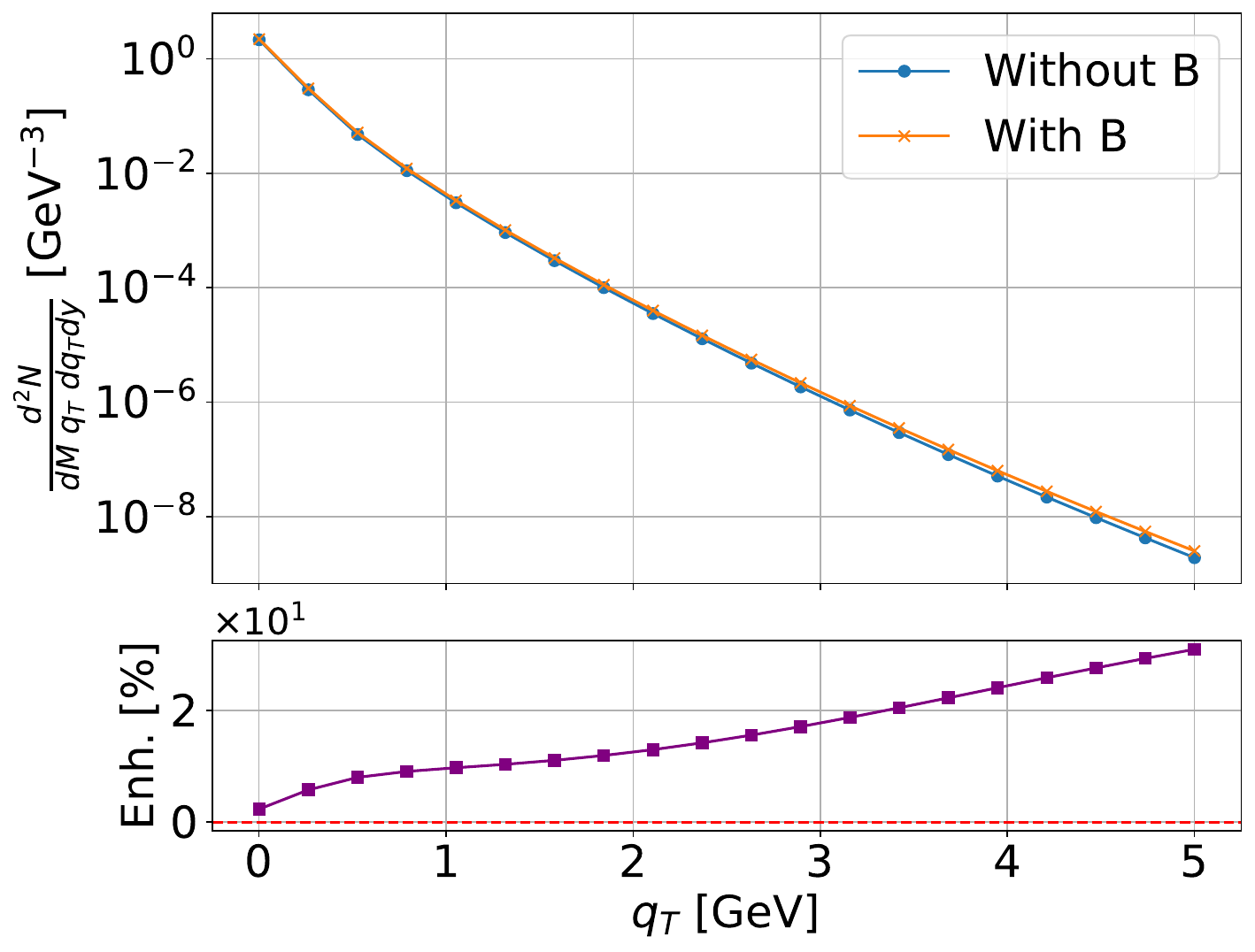}
	\includegraphics[width=0.4\textwidth]{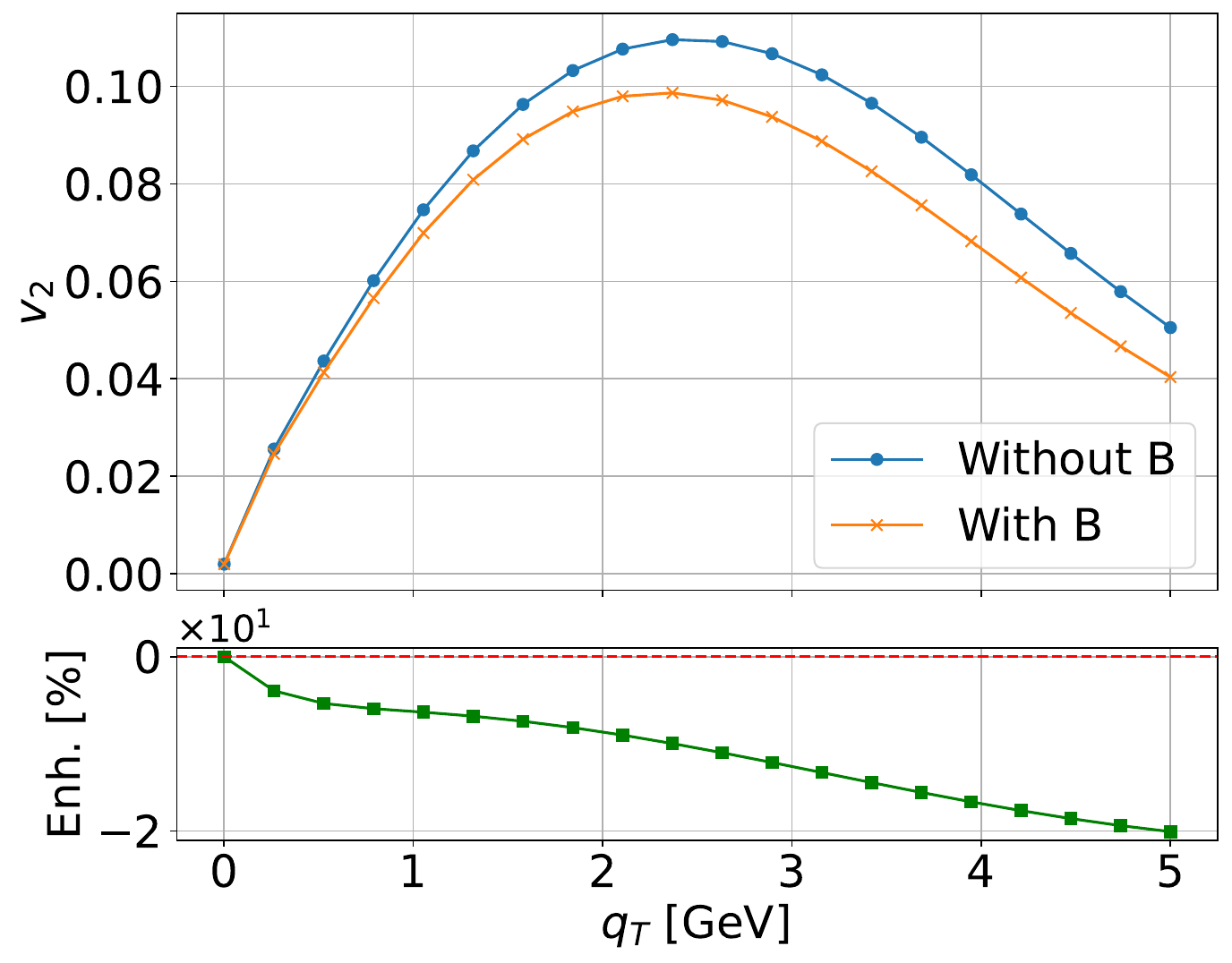}
	
	\vspace{0 cm} % Adds space between rows
	
	\includegraphics[width=0.4\textwidth]{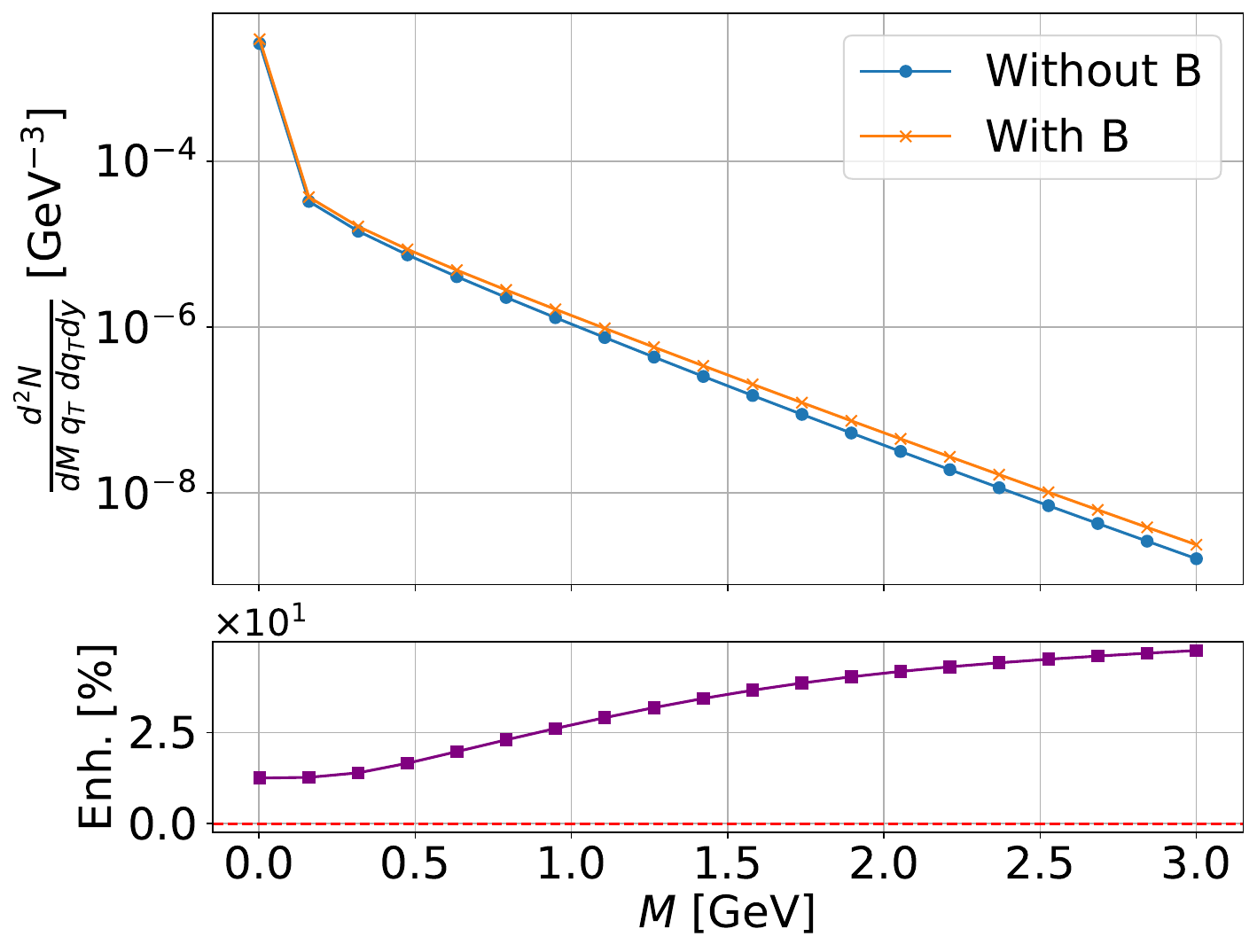}
	\includegraphics[width=0.4\textwidth]{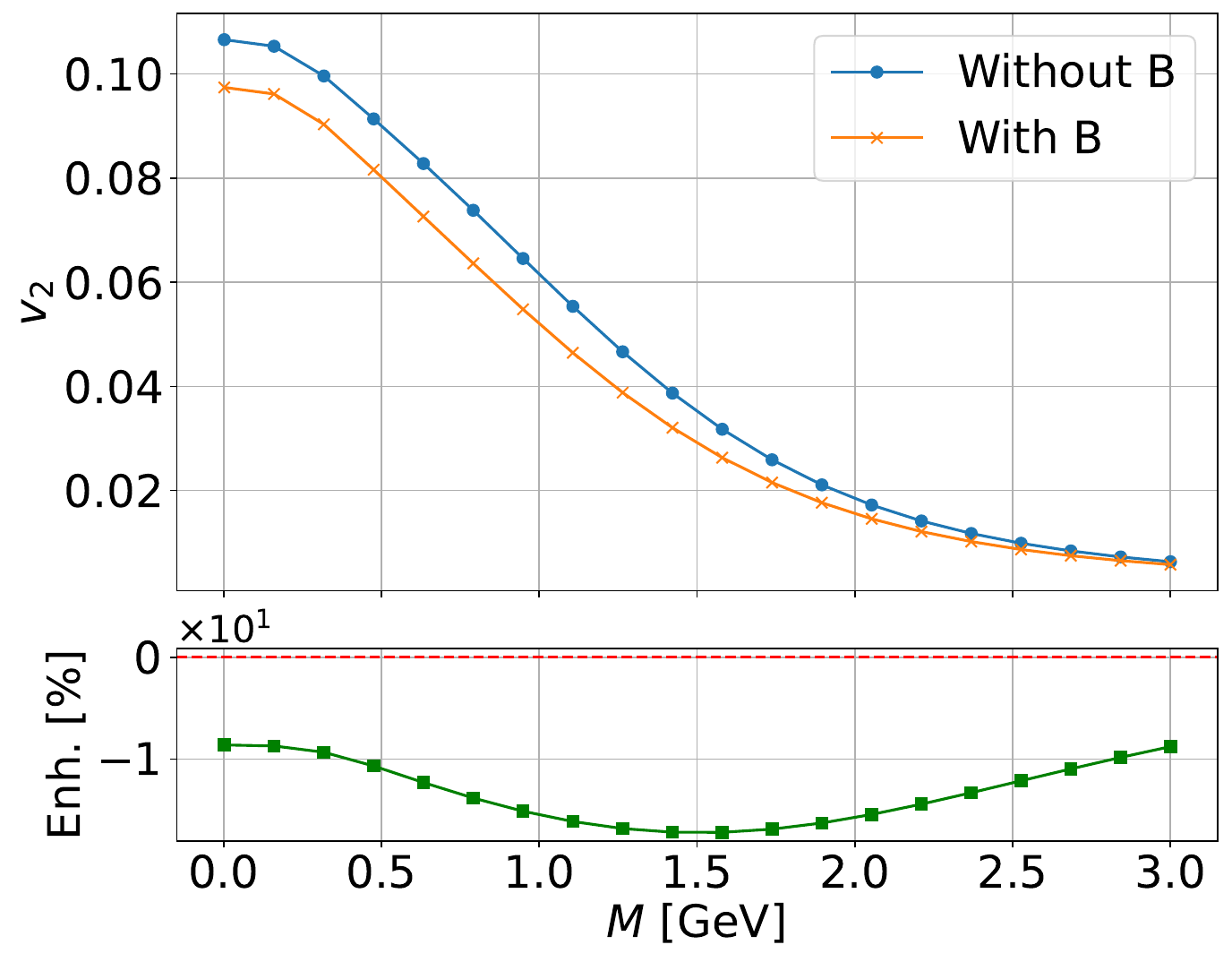}
	
	\caption{(Color online) Comparison of dilepton spectrum with and without magnetic fields with initial magnetic field strength of 10 $m_{\pi}^2$ are shown at mid rapidity. 
		Top panel contains the $q_{T}-$ spectra (left) with the variation $v_{2}(q_{T})$ (right) for fixed \( M = 0.1\,\mathrm{GeV} \). 
		Bottom panel depicts the $M-$ spectra (left) with the variation of $v_{2}(M)$ (right) for fixed \( q_T = 2\,\mathrm{GeV} \). 
		The relaxation time has been taken to be $\tau_{c}=5$ fm. The added lower sub-figures to each figure show percentage enhancement 
		$(v_{2}(B)-v_{2}(B=0))\times 100$ / $v_{2}(B=0)$.}
	\label{spectra_v_2_B10}
\end{figure*}

\begin{figure*}
	\centering
	\includegraphics[width=0.4\textwidth]{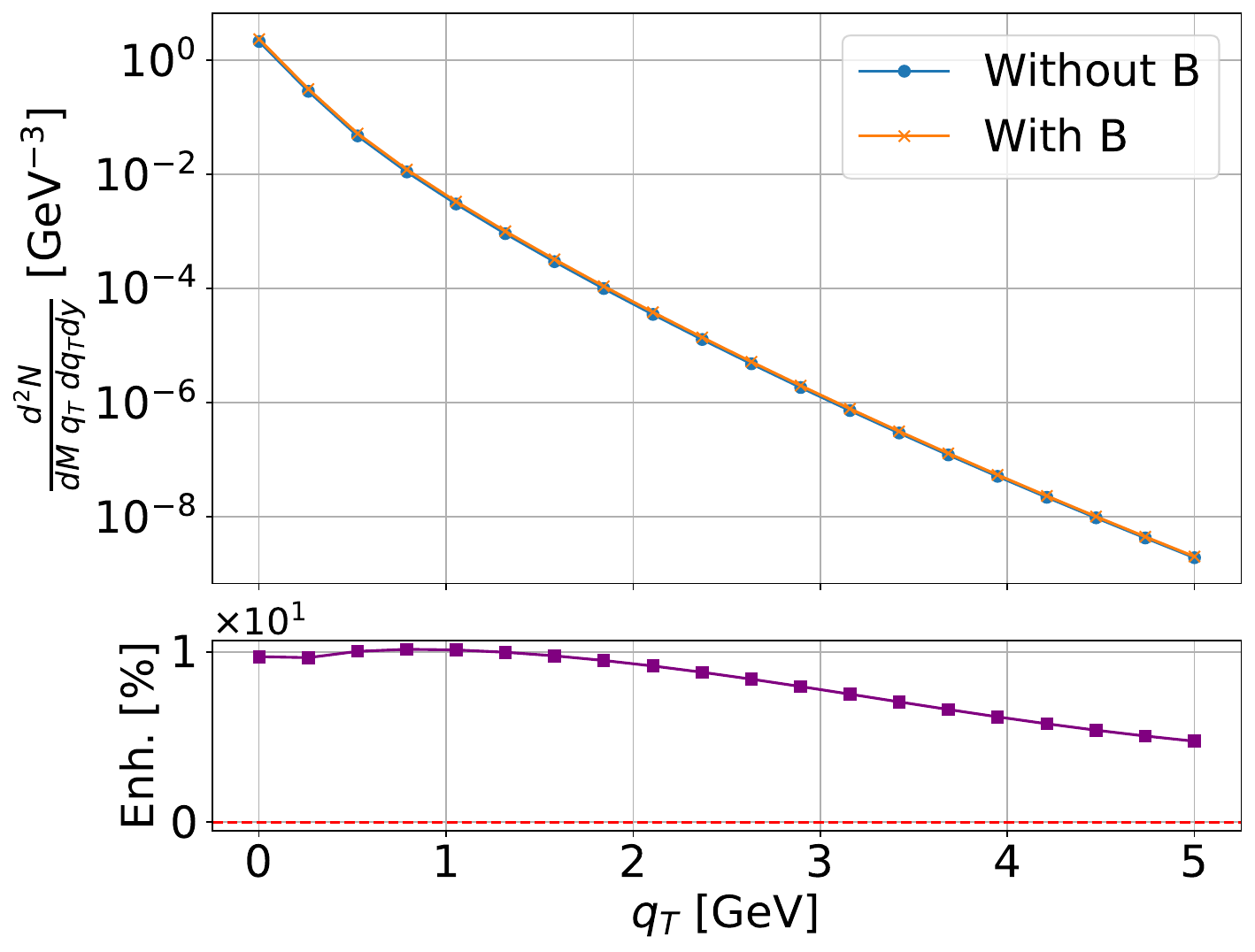}
	\includegraphics[width=0.4\textwidth]{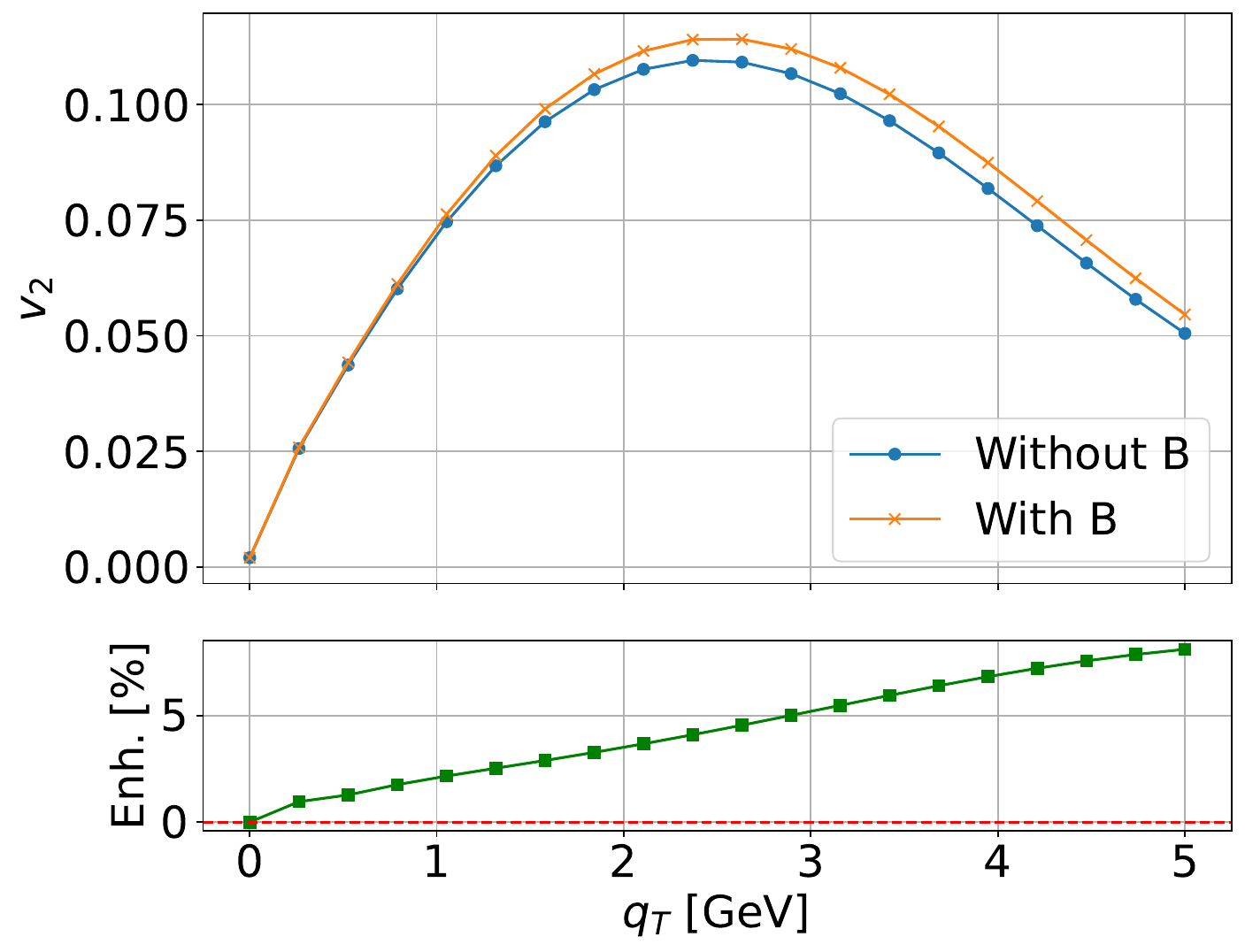}
	
	\vspace{0 cm} % Adds space between rows
	
	\includegraphics[width=0.4\textwidth]{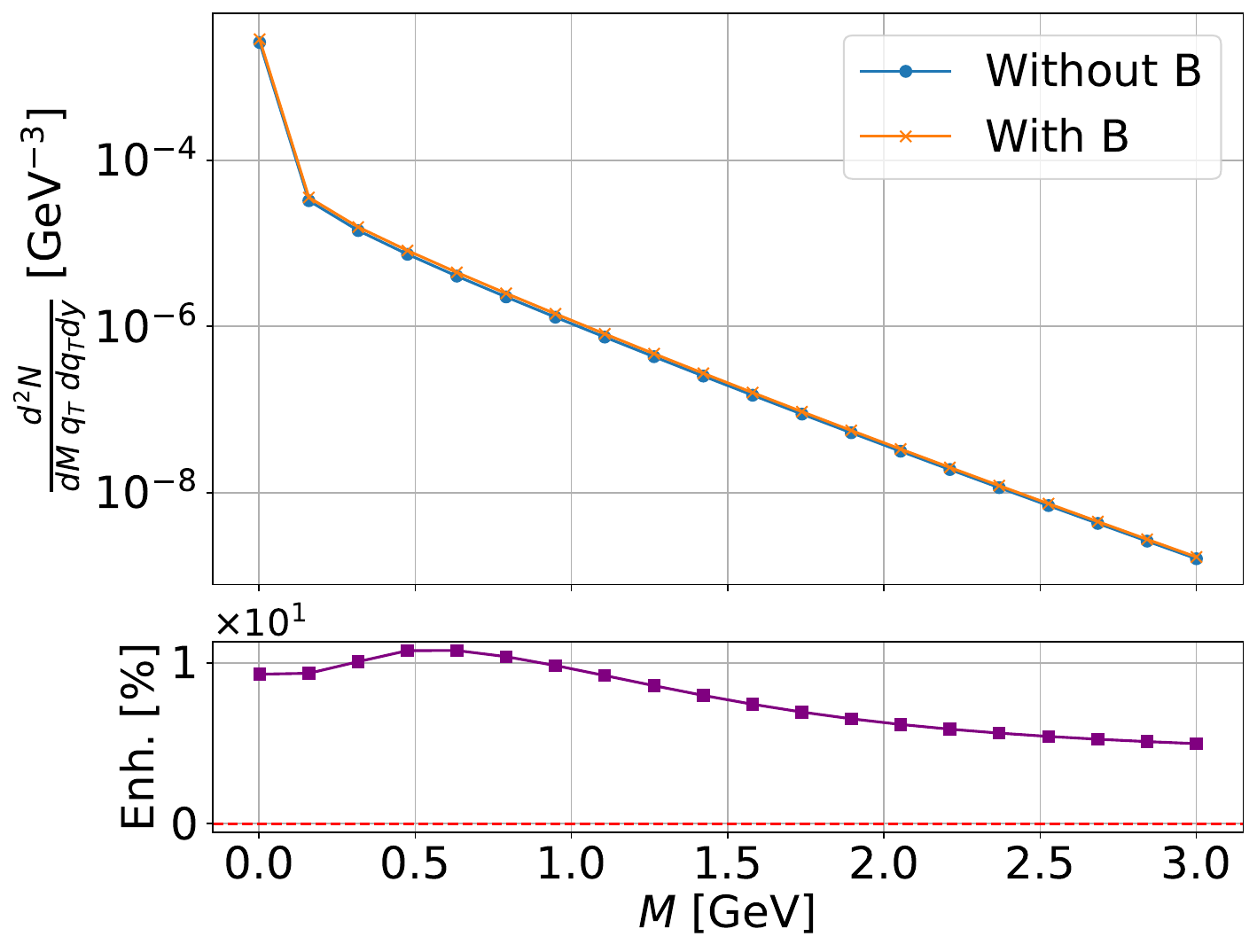}
	\includegraphics[width=0.4\textwidth]{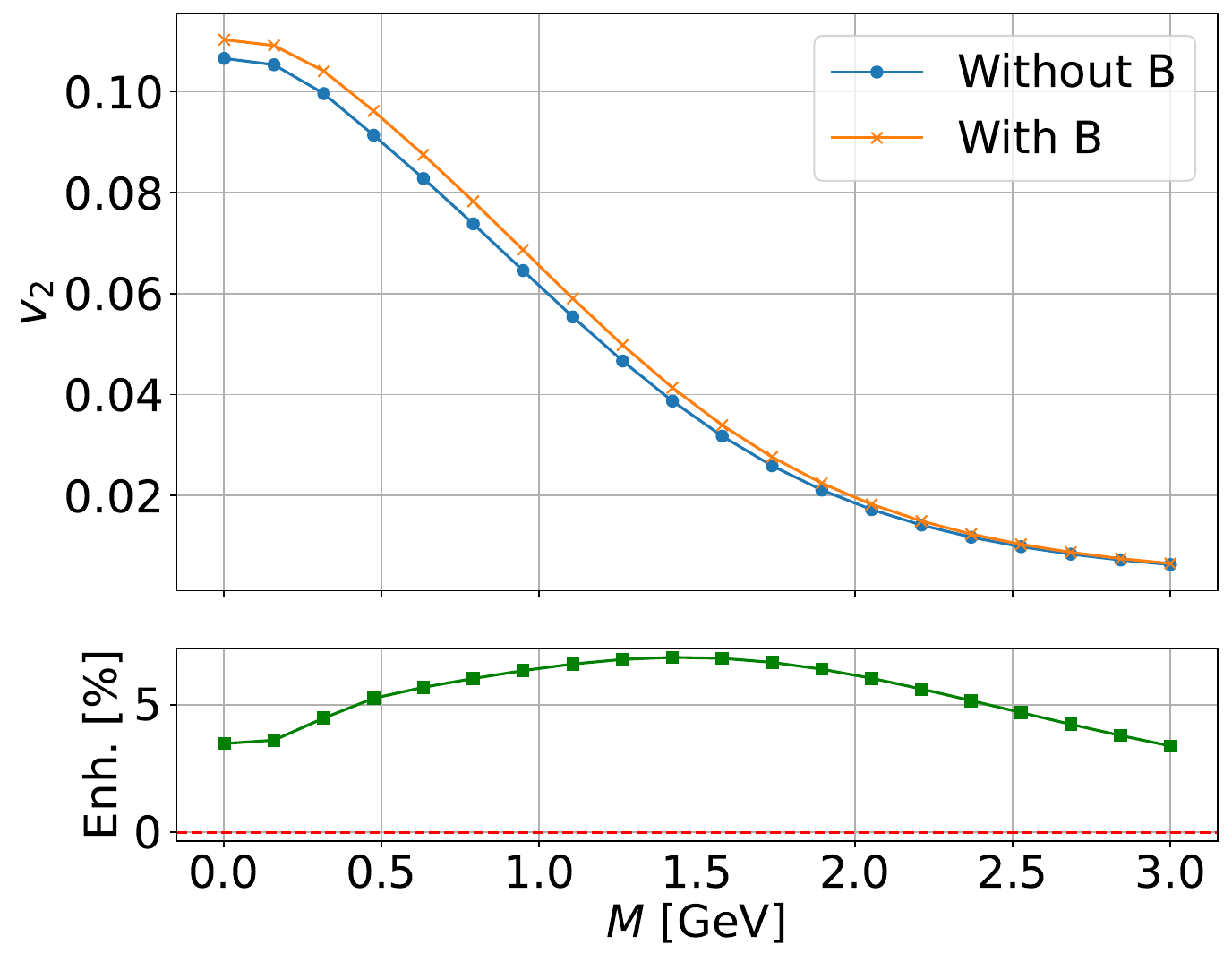}
	
	\caption{(Color online) Comparison of dilepton spectrum with and without magnetic fields for constant strength of 1 $m_{\pi}^2$ are shown at mid rapidity. 
		Top panel contains the $q_{T}-$ spectra (left) with the variation $v_{2}(q_{T})$ (right) for fixed \( M = 0.1\,\mathrm{GeV} \). 
		Bottom panel depicts the $M-$ spectra (left) with the variation of $v_{2}(M)$ (right) for fixed \( q_T = 2\,\mathrm{GeV} \). 
		The relaxation time has been taken to be $\tau_{c}=5$ fm. The added lower sub-figures to each figure show percentage enhancement 
		$(v_{2}(B)-v_{2}(B=0))\times 100$ / $v_{2}(B=0)$.}
	\label{spectra_v_2_BC}
\end{figure*}

After the discussion on the static rate and its variation with $T$, $\tau_{c}$ and $M$, we are all set to investigate the space-time integrated spectra and anisotropic flow using the fluid velocity and temperature profile coming from viscous hydrodynamics simulations (cf. Sec.~(\ref{Numerical setup})). The spectras and the elliptic flow are shown for two different values of $\tau_{c}=0.2$ fm (blue line) and $5$ fm (orange line) in Fig.~(\ref{spectra_v_2_noB}). For comparison, the corresponding QFT results by using the rate recorded in Eq.\eqref{QFTGhosh} are also displayed in green lines. In the top left panel of Fig.~(\ref{spectra_v_2_noB}) we delineate the variation of the dilepton transverse momentum spectra ($q_{T}-$spectra) at $M=0.1$ GeV for mid rapidity ($y=0$). For lower $q_{T}$ the spectra with the higher relaxation time (orange line) lies above the spectra with the lower relaxation time (blue line). For higher values of $q_{T}$ this trend revereses, i.e., $\tau_{c}=5$ fm curve lies below $\tau_{c}=0.2$ fm curve. This behavior can be understood by obtaining the behavior of the rate at two limiting cases as follows. For $M\ll|\vec{q}|$ the rate is proprtional to $\frac{1}{|\vec{q}|\tau_{c}}\ln [(2|\vec{q}|\tau_{c})^{2}+1]$ and for $|\vec{q}|\ll M$ the rate is proprtional to $\frac{M\tau_{c}}{M^{2}\tau_{c}^{2}+1}$. The QFT prediction aligns with the prediction of KTh at higher $q_{T}$ and for $\tau_{c}=5$ fm. Coming to the variation of the elliptic flow coefficients ($v_{2}$) as a function of $q_{T}$ we notice that it initially rises and then falls. The variation in $v_2(q_T)$ with different relaxation times shows relatively small differences across the $q_T$ range, whereas the $v_2(q_T)$ obtained from the QFT result remains consistently higher than the kinetic theory (KTh) result throughout the entire $q_T$ range shown. Next, in the bottom left and right panel of Fig.~(\ref{spectra_v_2_noB}) we respectively show the dilepton spectra and $v_{2}$ as a function of $M$. The $M-$spectra obtained at mid rapidity and constant $q_{T}=2$ GeV shows interesting characteristics. We observe that the curve for lower $\tau_{c}$ (blue line) lies above the higher $\tau_{c}$ results (orange line) while both of them show a decreasing trend with increasing $M$. The corresponding QFT results lie above the KTh curves for most of the part of the invariant mass $M$. The aforementioned difference between QFT and KTh results is easily reconciled by noticing the extra factor $\sim \frac{\tau_{c}}{M(M^{2}\tau_{c}^{2}+1)}$ for $M\gg |\vec{q}|$. Here, one can observe a rapid increase in the rate at lower $M$ for KTh which is orders of magnitude greater than the corresponding QFT results. This is expected as the ratio of the KTh static rate to QFT goes as $\frac{1}{M^{2}	|\vec{q}|\tau_{c}}\ln [(2|\vec{q}|\tau_{c})^{2}+1]$  for $ M/|\vec{q}|\ll 1$. Further looking at the variation of elliptic flow against the invariant mass we can observe an appreciable difference for different relaxation times at low $M$ with the difference decreasing with rising $M$. The $v_{2}(M)$ obtained from QFT lies above the KTh curve in the whole range of $M$ below 3 GeV as shown in bottom right panel of Fig.~(\ref{spectra_v_2_noB}).

We now move on from the discussion of dilepton spectra and flow in the absence of the magnetic fields and focus on how magnetic fields affects the spectra and flow harmonics. As we have discussed in the Sec.~(\ref{Conductivity calculation}) in the presence of magnetic fields the ratio of two time scales--cyclotron time period and relaxation time determine the impact of the magnetic fields on the dilepton production rate. In the present scenario, we firstly considered the peak magnetic field strength to be $1\ m_{\pi}^{2}$, attained at the center of the collision. The field decays with a power law in time, while exhibiting an exponential fall-off across the spatial volume. We checked that for space-time dependent magnetic field scenario $B_{y}=B_{y}(\tau,x,y)$ (cf. Fig.~(\ref{fig:By_contour})) the $M$ and $q_{T}$ spectra shows no appreciable change at a smaller relaxation time ($\tau_c$). In the left and right panel of Fig.~(\ref{fig:v2Bvary})
we respectively compare the $v_{2}(q_{T})$ at constant $M=0.1$ GeV and $v_{2}(M)$ at constant $q_{T}=2$ GeV for two different scenarios-- $B=0$ and $B_{y}=B_{y}(\tau,x,y)$. In this scenario, we find that the inclusion of magnetic fields results in negligible enhancement or suppression of the elliptic flow coefficient $v_{2}$ when compared to the case without magnetic fields. It is also checked that at a higher $\tau_c$ value of around 5 fm, the difference could be about 1 $\%$ level. When we increase the initial magnetic field strength from 1 $\,m_{\pi}^2$ to 10 $\,m_{\pi}^2$ keeping the relaxation time at around 5 fm, an appreciable change is observed in both spectra and $v_2$ as seen in Fig.\eqref{spectra_v_2_B10}.
Here we see that the changes in spectra is of the order of 20 $\%$ at both high $q_T$ and higher invariant mass, whereas a similar order of difference is also seen in $v_2$ but at an intermediate mass at a higher $q_T$ value.
Furthermore, under the presence of a constant magnetic field with strength $1\,m_{\pi}^{2}$ and a higher relaxation time of $\tau_{c} = 5$~fm, modifications of approximately $10\%$ in both the $q_{T}$ and $M$-spectra and a $5\%$ change in the elliptic flow is observed, as shown in Fig.~\eqref{spectra_v_2_BC}. The decaying magnetic fields showing a lesser suppression or enhancement than a constant valued field at a given relaxation time could be due to the fact that the magnetic field  which enters locally at each fluid cell through the effective relaxation time gets a smaller contribution due to its exponential fall off, which accumulate over the entire spacetime volume. In contrast, a constant magnetic field introduces a uniform and sustained modification across the entire volume, resulting in a more pronounced and coherent effect on the dilepton spectra and flow. An interesting and significant observation from Fig.~\eqref{spectra_v_2_BC} is the noticeable enhancement in both the $q_T$ spectra and elliptic flow at low invariant mass (around 0.1 GeV). This low-mass enhancement is also evident in the mass spectra and the $v_2$ as a function of mass, and can be attributed to the presence of the magnetic field—a phenomenon that has also been reported in earlier studies~\cite{Das:2021fma,Gao:2025prq}.

\section{ Summary and Conclusions  }\label{Conclusions and Summary}
To summarize, we have investigated thermal dilepton production from the quark-gluon plasma (QGP) created in heavy-ion collisions (HICs) at $\sqrt{s_{\rm NN}} = 200$~GeV. Starting from the general expression for the dilepton emission rate, we established its connection to the dynamical conductivity, which was derived within the framework of relativistic kinetic theory using the relaxation time approximation (RTA). This work presents a study of dilepton production rates that incorporates the relaxation time for quark–antiquark interactions. Here, we observed a non-monotonic behavior: the dilepton rate initially increases at lower $\tau_c$, reaches a maximum, and subsequently decreases. The position of this maximum is not universal; it shifts toward larger $\tau_c$ values for lower invariant masses, and vice versa.
When comparing our results with those obtained from traditional quantum field theory (QFT) approaches, we find clear differences. Specifically, for larger invariant masses, kinetic theory predicts lower yields than QFT, while for smaller masses, the kinetic theory rates are higher. These trends are reflected in both the transverse momentum ($q_T$) and invariant mass ($M$) spectra, obtained by integrating the local emission rates over the full space-time volume traversed by dileptons. The hydrodynamic evolution used for this integration was simulated using the MUSIC code at a fixed impact parameter of $b = 7$~fm.
We also examined the anisotropic emission pattern through the elliptic flow coefficient $v_2$ of dileptons. Our kinetic theory results consistently show smaller anisotropies compared to those from QFT. This trend was confirmed via azimuthal spectra studies (not included in the current manuscript), which demonstrate a systematically larger anisotropy in the QFT framework across the full mass range (up to 3~GeV) and transverse momentum range (up to 5~GeV).
So far, these analyses did not include magnetic field effects. To explore this aspect, we extended our framework by incorporating magnetic fields at the level of anisotropic conductivity—unlike some other approaches in the literature~\cite{Bandyopadhyay:2016fyd,Sadooghi:2016jyf,Bandyopadhyay:2017raf,Ghosh:2018xhh,Islam:2018sog,Das:2019nzv,Hattori:2020htm,Ghosh:2020xwp,Chaudhuri:2021skc,Das:2021fma,Wang:2022jxx,Mondal:2023vzx,Castano-Yepes:2024vlj,Gao:2025prq}. Two scenarios were considered: (i) a space-time dependent magnetic field model, and (ii) a simplified scenario with a constant magnetic field of strength $eB_y = 1\,m_\pi^2$ applied uniformly throughout the evolution.
We find that a lower initial magnetic field strength ($\sim 1 m_{\pi}^2$) with space-time variation has a negligible impact on both the dilepton spectra and elliptic flow—well below current experimental sensitivity at small $\tau_c$. However, at larger $\tau_c$ and higher initial field strengths ($\sim 10 m_{\pi}^2$), the results indicate noticeable effects. On the other hand, a constant magnetic field scenario with a higher $\tau_c$ value induces a modest $\sim$5-10\% modification in the bulk observables. The suppressed influence in case of space-time varying magnetic field at smaller relaxation times is attributed to its low magnitude across most of the space-time grid used in our simulation.

Beyond this current work, several related issues remain open for future investigation. 
An important extension would be to incorporate energy-dependent relaxation times into the kinetic theory framework, which could yield more realistic insights into quark–antiquark interaction dynamics. A more realistic scenario would require using a magnetohydrodynamic framework for QGP evolution when calculating dilepton rates. Additionally, the formalism developed here could be extended to study dilepton polarization, a topic of significant interest that may further enhance our understanding of early-time quark-gluon plasma dynamics. While these directions appear highly promising, they will require dedicated and detailed future studies.

\section{Acknowledgement}\label{Acknowledgement}
This work was partially supported by the Ministry of Education (MoE), Govt. of India (A.D.); IIT Bhilai Innovation and Technology Foundation (IBITF) (A.P.); and the Board of Research in Nuclear Sciences (BRNS) and Department of Atomic Energy (DAE), Govt. of India, under Grant No. 57/14/01/2024-BRNS/313 (S.G., V.R.).

\section{Appendix}
 In this appendix, we present three points: (1) the explicit expression for the trace of the retarded Green's function used in the dilepton rate, Eq.~(\ref{D4}); (2) the form of the dynamical conductivity in the limit $\omega, \vec{q} \to 0$; and (3) the expression of the dilepton rate in terms of the imaginary part of the dynamical conductivity.
 
 The trace of the dynamical conductivity for a given flavor can be written as:
 \bea
\Tilde{\sigma}^{k}_{k(f)}(\om, \vec{q})=-g~Q_{f}^{2} \beta  \int \frac{2d^{3}\vec{p}}{(2\pi)^{3}} \frac{ (v_{p})_{k}~(v_{p})^{k} f_{0}(1-f_{0})}{i(\vec{q}\cdot\vec{v}_{p}-\om)+\Gamma}.~\label{ape1}
 \eea
 Assuming the quarks to be massless, i.e., $v_{p}=1$ the above integral can be calculated analytically. Using the isotropy of the space we can fix $\vec{q}$ along the $z-$axis to evaluate Eq.~(\ref{ape1}) as,
 \bea
 \Tilde{\sigma}^{k}_{k(f)}(\om, \vec{q})=2g~Q_{f}^{2} \beta  \int \frac{d^{3}\vec{p}}{(2\pi)^{3}} \frac{f_{0}(1-f_{0})}{i(q\cos\theta-\om)+\Gamma}.\label{ape2}
 \eea
Using the result $-\frac{\del f_{0}}{\del \beta}=p~f_{0}(1-f_{0})$ and separating the real part and imaginary part we have,
 \bea
&&\Tilde{\sigma}^{k}_{k(f)}(\om, \vec{q})=-\frac{3\beta Q_{f}^{2}}{\pi^{2}}\int p\frac{\del f_{0}}{\del \beta}~dp\nn\\
&&\bigg[\int \frac{\Gamma~\sin\theta}{\Gamma^{2}+(\om-q\cos\theta)^{2}}+i\frac{(\om-q\cos\theta)\sin\theta}{\Gamma^{2}+(\om-q\cos\theta)^{2}}\bigg]d\theta,\nn 
 \eea
taking $\beta p=y$ for the integral outside the square bracket and $\om-q\cos\theta=x$ for the integrals inside the square bracket the above expression can be simplified to
\bea
\Tilde{\sigma}^{k}_{k(f)}(\om, \vec{q})&=&-\frac{3\beta Q_{f}^{2}}{\pi^{2}|\vec{q}|}\frac{\del}{\del \beta}\beta^{-2}\int_{0}^{\infty} \frac{y}{e^{y}+1}dy\nn\\
&&\bigg[\int_{\om-q}^{\om+q}\frac{\Gamma~dx}{\Gamma^{2}+x^{2}}+~i\int_{\om-q}^{\om+q}\frac{x~dx}{\Gamma^{2}+x^{2}}\bigg] \nn\\
&=&\frac{6T^{2}Q_{f}^{2}}{\pi^{2}|\vec{q}|}\eta(2)\left[\tan^{-1}\frac{x}{\Gamma}+i\ln \sqrt{\Gamma^{2}+x^{2}}\right]_{\om-q}^{\om+q}\nn
\eea
\bea
\implies \Tilde{\sigma}^{k}_{k(f)}(\om, \vec{q})&=&\frac{T^{2}Q_{f}^{2}}{2|\vec{q}|}\bigg[\tan^{-1}\left(\frac{2|\vec{q}|\Gamma}{\Gamma^{2}+M^{2}}\right)\nn\\
&&+\frac{i}{2}\ln \left(\frac{\Gamma^{2}+(\om+q)^{2}}{\Gamma^{2}+(\om-q)^{2}}\right)\bigg],\label{ape3}
\eea
 where to obtain the last line we used the standard law of combining logarithm and inverse tangent functions and used the result $\eta(2)=\frac{\pi^{2}}{12}$ ($\eta(j)$ is the Dirichlet eta function). From Eq.~(\ref{ape3}) we get the real part of the total conductivity as,
 \bea
&&{\rm Re}(\Tilde{\sigma}_{k}^{k}(\om,\vec{q}))=\sum_{f}\frac{Q_{f}^{2}T^{2}}{2|\vec{q}|}\tan^{-1}\left(\frac{2|\vec{q}|\Gamma}{\Gamma^{2}+M^{2}}\right). \label{ape4}
 \eea
 
From the famous Kubo relation one can get the DC conductivity $\sigma_{\rm DC}$ as a static limit of the retarded Green function~\cite{Fernandez-Fraile:2009eug},
\bea
\sigma_{\rm DC}&=&\frac{1}{3}\lim_{\om\xrightarrow{} 0^{+}}~\lim_{|\vec{q}|\xrightarrow{} 0^{+}}\frac{{\rm Im} G^{{\rm R} k}_{k}(\om,\vec{q})}{\om}\nn\\
&=& \frac{1}{3}\lim_{\om\xrightarrow{} 0^{+}}~\lim_{|\vec{q}|\xrightarrow{} 0^{+}}{\rm Re}\Tilde{\sigma}^{k}_{k}\nn\\
&=& \frac{1}{9}e^{2}T^{2} \lim_{\om\xrightarrow{} 0^{+}}~\lim_{|\vec{q}|\xrightarrow{} 0^{+}}\frac{1}{|\vec{q}|} \tan^{-1}\left(\frac{2|\vec{q}|\Gamma}{\Gamma^{2}+M^{2}}\right)\nn\\
\implies \sigma_{\rm DC} &=& \frac{2}{9}e^{2}T^{2}\tau_{c}\label{ape5}~,
\eea
where we have differentiated both the numerator and denominator to simplify the $\frac{0}{0}$ form inside the limit. 

The temporal component of the $\Tilde{G}^{{\rm (R)}\mu}_{\nu}$ enters the dilepton production rate. Using the relation $\tilde{G}^{{\rm (R)0}}_{0}=-\tilde{G}^{{\rm (R)0}}_{k}\frac{q^{k}}{\omega}$ and Eq.\eqref{D11} we obtain
\bea
\tilde{G}^{{\rm (R)0}}_{0}&=&2g\sum_{f}Q_{f}^{2} \beta  \int \frac{d^{3}\vec{p}}{(2\pi)^{3}} \frac{ i~(v_{p})_{k}~q^{k}f_{0}(1-f_{0})}{i(\vec{q}\cdot\vec{v}_{p}-\om)+\Gamma}\nn\\
&=&\frac{4g}{3}e^{2}\beta \int \frac{d^{3}\vec{p}}{(2\pi)^{3}} \frac{ (\vec{q}\cdot\vec{v}_{p})f_{0}(1-f_{0})}{(\omega-\vec{q}\cdot\vec{v}_{p})+i\Gamma}\nn\\
&=&\frac{e^{2}T^{2}}{3}\int \frac{ (\vec{q}\cdot\vec{v}_{p})}{(\omega-\vec{q}\cdot\vec{v}_{p})+i\Gamma}\sin \theta~ d\theta\nn\\
&=&\frac{e^{2}T^{2}}{3}\int \frac{q\cos\theta\sin\theta}{(\omega-q\cos\theta)+i\Gamma} ~d\theta\nn\\
&=& \frac{e^{2}T^{2}}{3}\bigg[\int \frac{q\cos\theta\sin\theta~(\omega-q\cos\theta)}{(\omega-q\cos\theta)^{2}+\Gamma^{2}}\nn\\
&&-i\Gamma\int \frac{q\cos\theta\sin\theta}{(\omega-q\cos\theta)^{2}+\Gamma^{2}}\bigg]~d\theta\nn
\eea
Upon simplification we get,
\bea
\tilde{G}^{{\rm (R)0}}_{0}&=&\bigg[\frac{1}{|\vec{q}|}\bigg(\frac{\omega}{2}\ln \frac{\Gamma^{2}+(\omega+|\vec{q}|)^{2}}{\Gamma^{2}+(\omega-|\vec{q}|)^{2}}+\Gamma \tan^{-1}\frac{2|\vec{q}|\Gamma}{\Gamma^{2}+M^{2}}-2|\vec{q}|\bigg)\nn\\
&-&\frac{i}{|\vec{q}|}\bigg( \omega\tan^{-1}\frac{2|\vec{q}|\Gamma}{\Gamma^{2}+M^{2}}-\frac{\Gamma}{2}\ln \frac{\Gamma^{2}+(\omega+|\vec{q}|)^{2}}{\Gamma^{2}+(\omega-|\vec{q}|)^{2}}\bigg)\bigg]\frac{e^{2}T^{2}}{3}\nn\label{DA1}\\
\eea
Using Eqs.~\eqref{ape4} and \eqref{DA1} we get the dilepton production rate as,
\bea
\frac{d^{4}N}{d^{4}x~d^{4}q}&=&\frac{\al}{12 \pi^{4}}~\frac{L(q^{2})}{q^{2}} ~f_{\rm BE} (\om) \left[ \om~ {\rm Re} \Tilde{\sigma}_{k}^{k}+{\rm Im} \Tilde{G}^{{\rm (R)}0}_{0}\right]\nn\\
&=& \frac{\al}{12 \pi^{4}}~\frac{L(q^{2})}{q^{2}} ~f_{\rm BE} (\om) \frac{e^{2}T^{2}}{6}\frac{\Gamma}{|\vec{q}|}\ln \frac{\Gamma^{2}+(\omega+|\vec{q}|)^{2}}{\Gamma^{2}+(\omega-|\vec{q}|)^{2}}\nn\\
&=& \frac{\al}{12 \pi^{4}}~\frac{L(q^{2})}{q^{2}} ~f_{\rm BE} (\om)3\sigma_{\rm DC} \frac{\Gamma^{2}}{4|\vec{q}|}\ln \frac{\Gamma^{2}+(\omega+|\vec{q}|)^{2}}{\Gamma^{2}+(\omega-|\vec{q}|)^{2}}\nn\label{DA2}\\
\eea
We observe that the above rate can be expressed as the imaginary part of the dynamical conductivity as
\bea
\frac{d^{4}N}{d^{4}x~d^{4}q}&=&\frac{\al}{12 \pi^{4}}~\frac{L(q^{2})}{q^{2}} ~f_{\rm BE} (\om)~\Gamma~ {\rm Im}(\sigma^{k}_{k})
\eea 
\bibliographystyle{unsrturl}
\bibliography{reference}
\end{document}